\documentclass{emulateapj}
\usepackage[latin1]{inputenc}
\usepackage{hyperref}

\shorttitle{Voronoi Tessellation cluster finder}
\shortauthors{Soares-Santos et al.}

\begin{document}

\title{The Voronoi Tessellation cluster finder in 2+1 dimensions}
\author{
  Marcelle Soares-Santos\altaffilmark{1,2}, 
  Reinaldo R. de Carvalho\altaffilmark{3},
  James Annis\altaffilmark{1},
  Roy R. Gal\altaffilmark{6},
  Francesco La Barbera\altaffilmark{5},
  Paulo A. A. Lopes\altaffilmark{4},
  Risa H. Wechsler\altaffilmark{7},
  Michael T. Busha\altaffilmark{7},
  Brian F. Gerke\altaffilmark{7}
}
\altaffiltext{1}{Fermi National Accelerator Laboratory, Batavia IL, USA;
marcelle@fnal.gov}
\altaffiltext{2}{Instituto de Astronomia, Geofísica e Ciências Atmosféricas,
  Universidade de São Paulo, São Paulo SP, Brazil}
\altaffiltext{3}{Divisão de Astrofísica, Instituto Nacional de Pesquisas 
  Espaciais, São José dos Campos SP, Brazil}
\altaffiltext{4}{Observatório do Valongo, Rio de Janeiro RJ, Brazil}
\altaffiltext{5}{INAF -- Osservatorio Astronomico di Capodimonte, 
Salita Moiariello 16, 80131 Napoli, Italy}
\altaffiltext{6}{Institute for Astronomy, University of Hawaii, 
  Honolulu HI, USA}
\altaffiltext{7}{Kavli Institute for Particle Astrophysics and Cosmology,
SLAC National Accelerator Laboratory, Stanford University, 
Stanford CA, USA}

\begin{abstract}
We present a detailed description of the Voronoi Tessellation (VT) cluster 
finder algorithm in 2+1 dimensions, which improves on past implementations of 
this technique.
The need for cluster finder algorithms able to produce reliable 
cluster catalogs up to redshift 1 or beyond and down to $10^{13.5}$ 
solar masses is paramount  especially in light of upcoming surveys aiming at 
cosmological constraints from galaxy cluster number counts.
We build the
VT in photometric redshift shells and use the
two-point correlation function of the galaxies in the field
to both determine the density
threshold for detection of cluster
candidates and to establish their significance.
This allows us to detect clusters in a self consistent way
without any assumptions 
about their astrophysical
properties. 
We apply the VT to mock catalogs which extend to redshift $1.4$ reproducing 
the $\Lambda$CDM
cosmology and the clustering properties observed in the SDSS data.
An objective estimate of the cluster selection function 
in terms of 
the completeness and purity as a function
of mass and redshift
is as important as having a reliable
cluster finder. We measure these quantities
by matching the VT cluster catalog with the mock truth table. 
We show that the VT can produce 
a cluster catalog with completeness and purity 
$>80\%$ for the redshift 
range up to $\sim 1$ and mass range down to $\sim 10^{13.5}$ solar masses.
\end{abstract}


\section{Introduction}

Today we recognize that galaxies constitute a very small
fraction of the total mass of a cluster, but they are nevertheless
some of the clearest signposts for detection of these massive
systems. Furthermore, the extensive evidence for differential
evolution between galaxies in clusters and the field -- and its
sensitivity to the underlying cosmological model -- means that it is
imperative to quantify the galactic content of clusters.
Perhaps even
more importantly, optical detection of galaxy clusters is now
inexpensive both financially and observationally. Large arrays of CCD
detectors on moderate telescopes can be utilized to perform all-sky
surveys with which we can detect clusters to $z\sim1$, and even
further with IR mosaics.

Forthcoming projects such as the Dark Energy Survey
(DES, \href{http://www.darkenergysurvey.org/}{\url{darkenergysurvey.org}}),
Pan-STaRRS
(\href{http://pan-starrs.ifa.hawaii.edu/}{\url{pan-starrs.ifa.hawaii.edu}})
and the Large Synoptic Survey Telescope
(LSST, \href{http://www.lsst.org/}{\url{lsst.org}})
will map thousands of square degrees to very faint limits
($\sim$29th magnitude per square arcsecond) in at least five filters,
allowing the detection of clusters through their weak lensing signal
as well as directly through the visible galaxies.
Combined with ever more efficient cluster-finding algorithms, these
programs will expand optical cluster detection to redshifts greater
than unity.
Prospects for utilization of such data to address one of the most
important scientific problems of our time by measuring the cosmological
parameters with improved precision are outstanding. In fact, given the
statistical power of these surveys, clusters have become
one of the strongest
probes for dark energy
\citep[e.g.,][]{Haiman:2001,Holder:2001,Levine:2002,Hu:2003,Rozo:2007b,Rozo:2010}\
.
Two unavoidable challenges imposed by these projects are to
produce optimal cluster catalogs -- with high completeness and
purity -- and to determine their selection function as a function of
cluster mass and redshift.

To see how to proceed, we must understand the strengths and important
limitations of techniques in use today, especially with respect to the
characterizability of the resulting catalogs.
We focus on photometric techniques rather than on
cluster finding in redshift space, which also has a long story,
starting with \citet{Huchra:1982}, and has been succesfully applied to
spectroscopic redshift survey data such as  2dFGRS \citep{Eke:2004}
and DEEP2 \citep{Gerke:2005}. Although the VT uses
redshift information, it is a photometric technique and this motivates
a discussion focused on this class of cluster finders.

The earliest surveys
relied on visual inspection of vast numbers of photographic plates,
usually by a single astronomer. The true pioneering work in this field
did not appear until the late fifties,
upon the publication of a catalog of galaxy
clusters produced by \citet{Abell:1958}, which remained the most
cited and utilized resource for both galaxy population and
cosmological studies with clusters for over forty years.
\citet*[hereafter ACO]{Abell:1989}
published an improved and
expanded catalog, now including the Southern sky. These catalogs have
been the foundation for many cosmological studies over the last
decades, even with serious concerns
about their reliability.
Despite
the numerical criteria laid out to define clusters in the Abell and
ACO catalogs, their reliance on the human eye and use of older
technology and a single filter led to various biases.
These old catalogs suffered as much from being black and white
as they did from being eye-selected.
Even more
disturbing, measures of completeness and contamination in the Abell
catalog disagree by factors of a few.
Unfortunately, some of these
problems will plague any optically selected cluster sample, but
the use of color information,
objective selection criteria and a strong statistical understanding of
the catalog can mitigate their effects.

Only in the past twenty years has it become possible to utilize the
objectivity of computational algorithms in the search for galaxy
clusters. These more modern studies required that plates be digitized,
so that the data are in machine readable form.
The hybrid technology of
digitized plate surveys blossomed into a cottage industry. The first
objective catalog produced was the Edinburgh/Durham Cluster Catalog
\citep[EDCC,][]{Lumsden:1992},
which covered 0.5 sr ($\sim 1,600$ square
degrees) around the South Galactic Pole (SGP). Later, the APM cluster
catalog \citep{Dalton:1997}
was created by applying Abell-like criteria to select
overdensities from the galaxy catalogs.
The largest, most recent, and the last of the
photo-digital cluster survey is the Northern Sky Optical Survey
\citep[NoSOCS; ][]{Gal:2000,Gal:2003,Gal:2009,Lopes:2004}.
This
survey relies on galaxy catalogs created from scans of the second
generation Palomar Sky Survey plates, input to an adaptive kernel
galaxy density mapping routine.  The final catalog covers 11,733
square degrees, with nearly 16,000 candidate clusters, extending to
$z\sim0.3$. A supplemental catalog up to $z\sim0.5$ was generated by
\citet{Lopes:2004} using Voronoi Tessellation and Adaptive Kernel maps.

With the advent of
CCDs, fully digital
imaging in astronomy became a reality. These detectors provided an
order-of-magnitude increase in sensitivity, linear response to light,
small pixel size, stability, and much easier calibration. The main
drawback relative to photographic plates was (and remains) their small
physical size, which permits only a small area (of order $15'$) to be
imaged by a larger $4096^2$ pixel detector. Realizing the vast
scientific potential of such a survey, an international collaboration
embarked on the Sloan Digital Sky Survey
(SDSS, \href{http://sdss.org/}{\url{sdss.org}}),
which included construction of a specialized 2.5 meter telescope, a
camera with a mosaic of 30 CCDs, a novel observing strategy, and
automated pipelines for survey operations and data processing. Main
survey operations were completed in the fall of 2005, with over 8,000
square degrees of the northern sky image in five filters to a depth of
$r'\sim22.2$ with calibration accurate to $\sim1-2\%$, as well as
spectroscopy of nearly one million objects.

With such a rich dataset, many groups both internal and external to
the SDSS collaboration have generated a variety of cluster catalogs,
from both the photometric and the spectroscopic catalogs, using
techniques including:
\begin{enumerate}
\item Voronoi Tessellation \citep{Kim:2002}
\item Overdensities in both spatial and color space
\citep[maxBCG,][]{Annis:1999,Koester:2007a,Hao:2009b}
\item Subdividing by color and making density maps
\citep[Cut-and-Enhance,][]{Goto:2002}
\item The Matched Filter and its variants \citep{Kim:2002}
\item Surface brightness enhancements
\citep{Zaritsky:1997,Zaritsky:2002,Bartelmann:2002}
\item Overdensities in position and color spaces, including redshifts
\citep[C4,][]{Miller:2005}
\item Friends-of-Friends \citep[FoF,][]{Berlind:2006}
\end{enumerate}
Each method
generates a different catalog, and early attempts to compare them have
shown not only that they are quite distinct, but also that
comparison of two photometrically-derived cluster catalogs, even from the same
galaxy catalog, is not straightforward
\citep{Bahcall:2003}.

In addition to the SDSS, smaller areas, but to much higher redshift, have
been covered by numerous deep CCD imaging surveys. Notable examples
include the Palomar Distant Cluster Survey
\citep[PDCS,][]{Postman:1996},
the ESO Imaging Survey
\citep[EIS,][]{Lobo:2000},
and many others.
None of these surveys provide the angular coverage
necessary for large-scale structure and precision cosmology studies, and
have been
specifically designed to find rich clusters at high redshift. The
largest such survey to date is the Red Sequence Cluster Survey
\citep[RCS,][]{Gladders:2005},
based on moderately deep two-band imaging using
the CFH12K mosaic camera on the CFHT 3.6m telescope, covers $\sim100$
square degrees. This area coverage is comparable to
X-ray surveys designed to detect clusters at $z\sim1$
\citep{Vikhlinin:2009b}.

Any cluster survey must make many different
mathematical and methodological choices.
Regardless of the data set and algorithm
used, a few simple rules should be followed to produce a catalog that
is useful for statistical studies of galaxy populations and for
cosmological tests:
\begin{enumerate}
\item Cluster detection should be performed by an objective, automated
algorithm to minimize human biases.
\item The algorithm utilized should impose minimal constraints on the
physical properties of the clusters, to avoid selection biases.
Any remaining biases must be properly characterized.
\item The sample selection function must be well-understood, in terms
of both completeness and purity, as a function of both redshift
and mass. The effects of varying the cluster model on the
determination of these functions must also be known.
\item The catalog should provide basic physical properties for all the
detected clusters, including estimates of their distances and some
mass proxy (richness, luminosity, overdensity) such that
specific subsamples can be selected for future study.
\end{enumerate}

One of the most popular and commonly used methods today
is the  Voronoi Tesselation
\citep[VT,][]{Ramella:2001,Kim:2002,Lopes:2004}.
Our implementation of this technique
is described in detail in \S \ref{algorithm}. Briefly,
it subdivides a spatial distribution into a unique set of polygonal cells,
one for each object, with the cell size inversely proportional to the local
density. One then defines a galaxy cluster as a high density region,
composed of small adjacent cells. Voronoi Tesselation satisfies the above
criteria for generating statistical, objective, cluster samples.
It requires {\it no} a priori assumption on galaxy colors, the presence of a
red sequence, a specific cluster profile or luminosity function.
Mock catalogs have been
used to test the efficiency of the detection algorithm.
These attractive qualities have led to its employment in numerous projects
beginning almost 20 years ago
\citep{vandeWeygaert:1989,Ikeuchi:1991,vandeWeygaert:1994,                      
Zaninetti:1995,El-Ad:1996,Doroshkevich:1997}.
\citet{Ebeling:1993}
used VT to identify X-ray sources as overdensities in X-ray photon counts.
\citet{Kim:2002}, \citet{Ramella:2001} and \citet{Lopes:2004}
looked for galaxy clusters using VT.
\citet{vanBreukelen:2009}
included the VT as one of two methods in their 2TecX detection algorithm,
an extension of their work on clusters in UKIDSS
\citep{vanBreukelen:2006}.
\citet{Barkhouse:2006}
used the VT to detect clusters
on optical images of X-ray Chandra fields.
\citet{Diehl:2006} applied a modified version of the VT algorithm
to X-ray data.

Here we improve on past implementations of this technique focusing on
optical data. We build the
VT in photometric redshift shells and use the
two-point correlation function of the galaxies in the field
to determine the density
threshold for detection of cluster
candidates and to establish their significance.
This allows us to detect clusters in a self consistent way using a minimum
set of free parameters and without any assumptions  about the astrophysical
properties of the clusters. We provide a list of member galaxies for
each cluster and use the number of members as a proxy for mass.
We apply the VT on mock catalogs that accurately reproduce the $\Lambda$CDM
cosmology and the clustering properties observed in the SDSS data.
By comparing the VT cluster catalog with the truth table, we
measure the completeness and purity of our cluster catalog as a function
of mass and redshift. We show that our implementation of the VT produces
a reliable cluster catalog up to redshift $\sim 1$ and down to
$\sim 10^{13.5}$ solar masses.

The paper is organized as follows: \S\ref{algorithm} is dedicated to
a detailed presentation of the algorithm; \S\ref{efficiency} describes
the method used to compute the selection function  of the cluster
catalog; in \S\ref{results} we discuss the completeness and purity
results and show our ability to recover the mass function of the mock
catalog at redshift close to unity; \S\ref{summary} presents a summary
of this work. The work on the relation between the two-point correlation
function and the VT cell areas distribution -- fundamental for the
development of our method -- is detailed in the Appendix.

\section{Algorithm}\label{algorithm}
We present the VT cluster finder in 2+1 dimensions. The method is 
non-parametric and does not smooth the data, making the detection independent
of the cluster shape. It uses all  galaxies available, going as far down
in the luminosity function as the input catalog permits.
It does not rely on the existence of features such as a unique 
brightest cluster galaxy or a tight ridgeline in the
color-magnitude space. 
It works in shells of redshift,
treating each shell as an independent 2-dimensional field. 

Central to the VT algorithm is the background over which an overdensity 
must rise to be identified as a cluster. In contrast to earlier 
implementations of the 
VT algorithm \citep{Ebeling:1993,Ramella:2001,Kim:2002,Lopes:2004}, 
we do not assume a Poissonian background. We use a more 
realistic assumption that the angular two-point correlation function 
of the background galaxy distribution is represented by  a power-law
\citep[e.g.][]{Connolly:2002}.
Another improvement over earlier works on VT-based cluster finders
is the use of photometric 
redshifts instead of magnitudes \citep{Ramella:2001,Lopes:2004} or
colors \citep{Kim:2002}. This eliminates the need for a percolation 
step and allows for a cluster finder which is not 
based on astrophysical properties of clusters (the luminosity function or 
color-magnitude relation), but on the characteristics 
of the large scale clustering process. This makes the VT a 
cluster finder subject to different systematics from color-based 
methods.
  
The fundamental inputs required for cluster detection using the VT are 
the coordinates RA, Dec 
and redshift of each galaxy and the redshift error $\sigma_z(z)$ 
for the full galaxy sample.
The input catalog is sliced in non-overlapping 
1-$\sigma_z$ wide redshift shells. 
Note that  the velocity dispersion of a typical cluster is 
much smaller than realistic values of $\sigma_z$.
For each shell an estimate of the 
parameters ($A$,$\gamma$) of the two-point correlation function 
is required. This can be obtained directly from the data.
  
We then build a Voronoi diagram and compare the 
distribution of cell areas with the distribution expected from a 
background-dominated field. 
Since small cell size implies high density, this allows us to 
establish a  size threshold below which 
the distribution is dominated by cluster members. 
The most significant clumps of contiguous cells smaller than this threshold 
are listed as clusters. This procedure is repeated on all 
redshift shells and the results are merged into a unique list of 
cluster candidates. The merge proceeds as follows.
From the input galaxy catalog we  
extract 3-dimensional boxes centered at the coordinates of each candidate.
We run the VT on those boxes to confirm the 
detection. This recursive procedure eliminates the edge effects at the
interface between successive shells, reduces the number of fake detections
due to projection effects and eliminates multiple detections.

In the resulting cluster catalog, we report position, redshift, redshift 
error, galaxy density contrast, significance of 
detection, richness, size and shape parameters of the clusters. 
We also provide a list of members
with the local density of their 
respective cells and flags indicating the central
galaxy (the galaxy found in the highest density cell). 

Although it is possible to build Voronoi diagrams on a sphere, we use 
a rectangular coordinate system, which is easier to implement.
This implies that we must process small sky areas at a time to 
avoid distortions due to tangential projection. 
We have tested 
different area sizes and concluded that boxes of $3\times3$ degrees
are adequate.
A buffer region is implemented to avoid edge effects and the
effective area is the central $1 \times 1$ square degree box. 
Clusters found in the buffer regions are rejected prior to 
the merging of the shells' candidate lists.
The size of the buffer zone corresponds to the angular 
scale of a large cluster at the lowest redshift  
(a 1 degree scale corresponds to $\sim3$ Mpc at $z=0.05$).  


In the following, we detail each step of the cluster detection 
process and explain
how each of the above quantities are derived, 
justifying the choices made in designing the algorithm. 

\subsection{VT construction}\label{vtconstruction}
\begin{figure}[ht]
\begin{center}
\includegraphics[width=75mm]{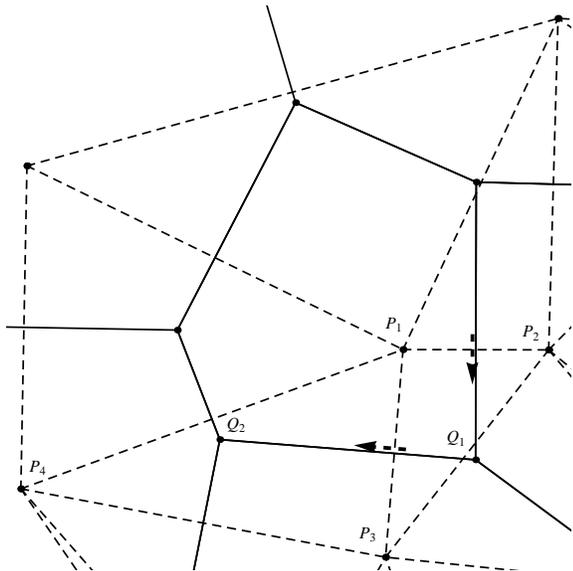}
\end{center}
\caption{A portion of a typical Voronoi tessellation is shown
together with its dual Delaunay mesh (solid and dashed lines, respectively) 
to illustrate the Voronoi diagram building process.
For each generator set $P_i$, there is one and only one set of 
Voronoi cells given by the vertices $Q_i$. See text for details.} 
\label{building_vt}
\end{figure}

The Voronoi diagram of a 2-dimensional distribution 
of points is a unique, non-arbitrary and 
non-parametric fragmentation of the 
area into polygons. A simple algorithm to perform such fragmentation 
is the following (see Fig.~\ref{building_vt}): 
starting from any position
$P_1$, we label its nearest neighbor $P_2$ and walk 
along the perpendicular bisector between those points.  
We stop when we reach for the first time a point $Q_1$ equidistant 
from $P_1$, $P_2$ 
and any third point $P_3$. 
We now walk along the perpendicular bisector between
$P_1$ and $P_3$ until we reach the point $Q_2$ and identify the next
point $P_4$ by the same criterion. Successive repetition of this process
will eventually bring us back to $Q_1$ after a finite number of steps. 
The set of points $Q_i$ are the vertices  of a polygon, the Voronoi cell, 
associated with $P_1$. If this process is repeated for each point $P_i$ 
we will have built the Voronoi tessellation corresponding to this point field.

However, there are several more 
robust and efficient computational algorithms to 
build a Voronoi diagram from a given distribution. In our code we
use the so-called Divide \& Conquer algorithm implemented in the Triangle
library \citep{Shewchuk:1996}. The D\&C is based on recursive partition
and local triangulation of the points and then on a merging stage. The 
total running time, for a set of $n$ points is $O(n\log n)$.

There are no arbitrary choices in building the VT.
The cell edges are segments of the perpendicular bisectors between 
neighbor points and each vertex is an intersection 
of two bisectors. This implies that the cells will be smaller in the 
high-density regions
and since each cell contains one and only one point, the inverse of the 
cell area gives the local density. The VT cluster finder takes 
advantage of this fact in the process of detection.  

\subsection{Cluster candidate detection}
Each realization of a given point process will result in a distinct 
unique tessellation, but the distribution of Voronoi cell areas will be
the same.  The case of the Poisson point process has been extensively 
investigated and it has been shown \citep{Kiang:1966} that the resulting 
distribution of Voronoi cell areas is well fitted by a gamma distribution
\begin{equation}\label{gamma_distribution} 
p(x) = \frac{\beta^{\alpha}}{\Gamma(\alpha)} x^{\alpha -1 } \exp^{-\beta x}
\end{equation}
with $\beta=\alpha=4$ (only for the Poisson case) 
and $x$ being the cell area normalized by the mean 
area of all cells. Here we extend Kiang's formula to a more general case.

Consider a random distribution of points in a plane with two-point 
correlation function given by $w(\theta)=A \theta^{1-\gamma}$, where
the variable $\theta$ is the separation between point pairs and the
parameters  $A$ and $\gamma$ are respectively the amplitude and slope of the 
power-law. The Poisson distribution is the  particular case 
where  $A \to 0$.
A general relation between 
the statistics of the point field and the VT areas distribution remains 
as a conjecture yet to be proved, but  
in the case of a point field 
generated from the above two-point correlation function, 
the gamma distribution still holds with the  
values of $\alpha$ and $\beta$ modified.
We have proven this fact and obtained the relation between $\alpha,\beta$ 
and the parameters $A,\gamma$ numerically. Using the 
simulated annealing method  described in the context of materials 
science \citep{Rintoul:1997} we generate test fields  spanning a 
wide range of $A,\gamma$ pairs. On each test field we applied the VT algorithm 
and obtained the corresponding distribution of cell areas, fitting 
Eq.~\ref{gamma_distribution} to obtain the corresponding pair 
$\alpha,\beta$. These two parameters are not independent. They are
related by a simple relation: $\beta = \alpha - 0.26$. 
See the Appendix for a detailed discussion of these results.

Information about the background is given to the VT code via the two input  
parameters $A,\gamma$. These will depend on the redshift shell and, 
ideally, they should
be estimated directly from the data being considered. 
High accuracy in
the parameters are not required, though. 
Note that no free parameters are introduced by $A$ and $\gamma$, since they
can be completely determined from the global input galaxy catalog. 
Clusters and groups present in the field when the 
two-point correlation function is measured do not affect 
the cluster finder. On the contrary, our method 
is based on the idea that the clustering process resulting in 
the power-law described by $A$ and $\gamma$ also results in 
the formation of clusters, which are found in the high density 
end of the VT cell distribution.

\begin{figure}[ht]
\begin{center}
\includegraphics[width=75mm]{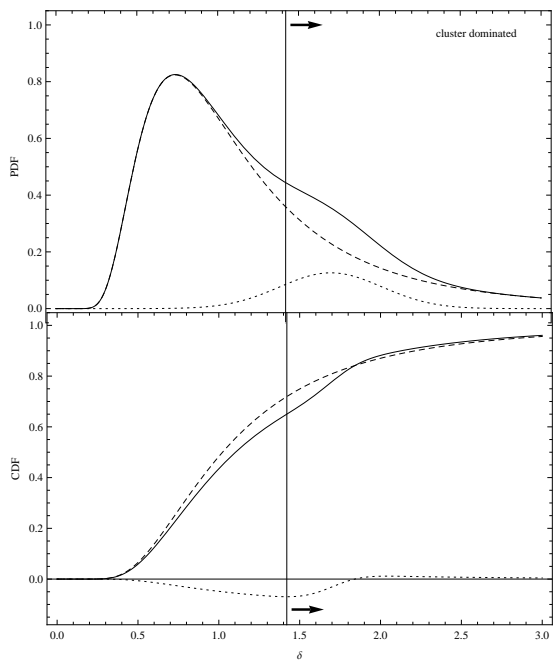}
\end{center}
\caption{Differential and cumulative distributions of normalized cell 
densities illustrating the process of detection in the VT cluster finder.
The dashed lines correspond to a background distribution with $A=0.005$
and $\gamma=1.7$. The solid lines correspond to the distributions 
distorted by an artificial Gaussian-shaped cluster contribution (dotted line).
The vertical line is the threshold for detection $\delta^*$. All cells above 
the threshold are selected as cluster member candidates.
}
\label{pdf_cdf_illustr}
\end{figure}

Taking the differential probability distribution  
(\ref{gamma_distribution})  as a function of the normalized cell density, 
$\delta = 1/x$, our goal is to identify a density threshold $\delta^*$
above which the contribution of the clusters starts to dominate over the 
background. A schematic example is shown in 
Fig.~\ref{pdf_cdf_illustr}. To the background 
distribution given by $A=0.005$ and $\gamma=1.7$ (upper panel, dashed line), 
we add a 
cluster contribution of $10\%$ given by a simple Gaussian (upper panel,
dotted line). As a result, the total distribution is distorted by the 
presence of the clusters.
To perform the detection, we take the corresponding  cumulative 
distributions. For the background, the cumulative distribution is given by, 
\begin{equation}\label{gamma_distribution_cum}
P(\delta) = \frac{\Gamma (\alpha, \beta/\delta)}{\Gamma (\alpha)} 
\end{equation}
and depends on the input parameters $A,\gamma$ through $\alpha$ and $\beta$. 
The maximum of the difference between the background (dashed) and the total 
(solid) distributions corresponds to the point where the total distribution
increases faster than the background. This point is a natural choice for 
the threshold $\delta^*$ (vertical line). 
 
In the example above an artificial cluster contribution with a particular
shape was added  to illustrate the   principle of detection. In the 
actual process, we work only with the cumulative distributions. 
Once the threshold is computed we select all the cells with 
$\delta \geq \delta^*$.
We then take the clumps of contiguous selected cells as cluster
candidates.

Setting  the threshold at the point of maximum difference between the 
two distributions leads to the detection only of the central regions of 
the most massive clusters ($M>10^{14.5}M_{\odot}$). 
This is a consequence of the fact that the two-point correlation function
of the field includes the contribution of clusters, and only the 
highest density peaks deviate significantly from the distribution
predicted by Eq. \ref{gamma_distribution_cum}.
To improve this 
result, we allow this to be an adjustable parameter, called $scl$.
By comparing the two-point correlation function
of galaxies measured by \citet{Davis:1983} in the $14.5 m_B$ 
CfA redshift survey with the two-point correlation function of   
rich ($R\geq1$) Abell clusters measured by \citet{Bahcall:1983},
\citet{Bahcall:1986}  has estimated 
that $\sim 25\%$ of all galaxies are associated with clusters and 
the  10 Mpc scale structures that surround them. We therefore set 
our threshold at the point $\delta^*$ where the cumulative 
distribution reaches $\sim 75\%$. 
As this fraction must change with redshift, magnitude limit of the
galaxy catalog and lower mass limit of the cluster catalog, we
determine   
the exact values of the cumulative distribution used to set $\delta^*$ 
in each redshift bin, $scl(z)$,  
by applying the cluster finder on simulated galaxy catalogs and 
maximizing the completeness and purity of the output  catalog.  
This process does introduce a free parameter that we must tune.

\subsection{Selection of high-significance candidates and membership assignment}

For a given threshold $\delta^*$, we assume that 
each cluster candidate has a probability
\begin{equation}\label{clump_probability}
p(\delta_{min},N_g)=1-{\mathrm{Erf}} \left(\left(\frac{\delta_{min}}{\delta^*}-1\right)\frac{N_g}{\sqrt 2} \right)
\end{equation}  
of being caused by random fluctuations of the background field. 
Here $\delta_{min}$ is the minimum cell density and $N_g$ is the number of
galaxies in the candidate.
Note that the process of detection implies $\delta_{min} \geq \delta^*$.
A confidence level of $95\%$ is required for a candidate to be accepted. 
If a given candidate has $p(\delta_{min},N_g)$ below this level, we 
iterate on its cells, dropping the one with lowest density and recomputing
$p(\delta_{min},N_g)$, until this candidate falls within the acceptable level
or runs out of galaxies. 
As a result, some cluster candidates will be reduced
in size and others will be eliminated. 
The final list of candidates is composed of clusters above the required
confidence level. 
This cleaning  process 
is  necessary as the  $\delta^*$ thresold is set to be permissive;
the estimate by \citet{Bahcall:1986} that $\sim25\%$  of all 
galaxies are associated with clusters was accompanied by a 
hypothesis that these galaxies were distributed 
in $\sim30$ Mpc scale overdense regions about clusters, while
we aim to detect clusters closer to the $\sim1Mpc$ Virial
scale. 
This process results in a list of cluster members, given 
by all the galaxies within the final VT footprint of the cluster. 
The galaxy belonging to the cell of highest density is taken as the central 
galaxy. 

The accuracy of the membership assignment 
is limited by the errors in the redshift of the galaxies and width of the
redshift shell. As discussed in section \ref{catalog_construction}, 
the membership list is improved in the second run 
of the VT cluster finder, which is performed in boxes 
centered at the central galaxies flagged 
during this first run. 

\subsection{Shape measurement}\label{shapemeasurement}
To obtain the cluster shape parameters, we take the galaxies within the
cluster VT footprint and compute the second moments of 
the galaxy distribution with 
respect to the coordinates ($x_c,y_c$) of the central galaxy, using the cell 
densities $\delta$ as weights.
These second moments are:
\begin{eqnarray}
 m_{xx}= \frac{\sum_i \delta_i (x_i-x_c)^2}{\sum_i \delta_i} \nonumber \\
 m_{yy}= \frac{\sum_i \delta_i (y_i-y_c)^2}{\sum_i \delta_i}\\
 m_{xy}= \frac{\sum_i \delta_i (x_i-x_c)(y_i-y_c)}{\sum_i \delta_i} \nonumber
\end{eqnarray}
where the $x$ and $y$ directions are aligned with the RA and Dec axes,
respectively. 
We use these quantities to compute the semi-major and semi-minor axes, 
$a$ and $b$, respectively:
\begin{eqnarray}
a=\left[\frac{1}{2}(m_{xx}+m_{yy}+f)\right]^{1/2} \nonumber \\
b=\left[\frac{1}{2}(m_{xx}+m_{yy}-f)\right]^{1/2}
\end{eqnarray}
where $$f=(m_{xx}-m_{yy}+4m_{xy})^{1/2}$$
The position angle is also obtained in terms of the same quantities,
\begin{equation}
PA=\frac{180}{\pi} \tan^{-1} \left(\frac{b^2 -m_{xx}}{m_{xy}}\right)
\end{equation}
and is given in degrees.

\subsection{Catalog construction}\label{catalog_construction}
A global list of cluster candidates is made by merging the results
of the individual shells. For each cluster in that list we extract from
the full input galaxy catalog (not the $z$ shells) a 
3-dimensional box centered at its central galaxy and with the 
same size as in the first run: $3\times3$ square 
degrees and $\sigma_z$ width. 
These boxes are processed with the VT algorithm, repeating 
the steps described in \S\ref{vtconstruction}-\ref{shapemeasurement}
 and a new global list of
cluster candidates is constructed, taking only the clusters found at the 
center of each box.
 
We perform a  matching between the two global lists. In this matching 
scheme, candidates are considered the same cluster if they have more than 
50\% of shared galaxies and multiple matches are not allowed. When a 
matching occurs, that cluster is eliminated from the list of 
candidates available for matching with other candidates.
The clusters
found in the first run but undetected in the second run are eliminated as
projection effects. The primary function of this stage, however, is to 
deal with photo-z slice edge effects.

Because the new boxes are allowed to cross the initial shell boundaries, 
edge effects in the redshift dimension are eliminated. 
Clusters split in 
several components during  the initial detection will result in 
cluster candidates with a number of shared galaxies after the second run.
For a given pair of candidates found to be the same cluster 
(i.e., sharing more than 50\% of their galaxies),
only the one with the largest number of 
members is added to the final cluster catalog. 
Otherwise, they are said to be 
distinct clusters with shared galaxies (which are flagged in the members
list) and both are included in the cluster catalog.
Setting the threshold of shared galaxies to 50\% is a 
natural choice between the two extremes where all candidates 
would be duplicated or only the clusters found with the same 
set of member galaxies would be accepted. 

At this point the detection is completed. 
We have the final list of clusters containing RA, Dec, redshift 
and a list of member galaxies including the parameters 
of the corresponding VT cells. This forms the VT footprint of the cluster.
The cluster redshift is estimated as the median of the redshift of the 
cluster members. 
The quantity is better estimated in the second run after a cleaner 
membership list is obtained, so as to avoid projection effects along the
line of sight.

The output parameters of the VT cluster catalog are:
ID, RA, Dec (coordinates of its central galaxy or the highest density peak), 
z (given by the median of all members), $\sigma_z$ (rms value), 
$\delta_c$ (density contrast measured at the final stage of detection), 
$\sigma$ (significance of detection), richness (number of members), 
size (radius of the circle enclosing all galaxies), a (semi-major axis), 
b (semi-minor axis) and PA (position angle).

We also report a members list containing: 
ID, host ID (most likely host cluster), cell density, 
shared flag (1 if the galaxy is shared with another cluster, 
0 otherwise) and
central flag (1 for central galaxy, 0 for regular 
members).
Note that we do not list every possible galaxy-cluster association in 
the output.
Galaxies not associated to any cluster are listed 
with host ID, shared flag and central flag set to $-1$. 
These non-member galaxies can be 
used, for instance, to compute the local density of non-member galaxies 
around a cluster or to run afterburners to measure cluster properties 
such as richness and $R_{200}$.

Having a list of members generated by the cluster finder is 
highly desirable, because properties such as the optical richness and 
$R_{200}$ can be estimated. The lack of membership assignment in
VT implementations using magnitudes was a drawback and we improve on that
matter. 
Also, this 
allows us to compute the algorithm efficiency as follows.

\section{Algorithm efficiency}\label{efficiency}
The effectiveness of the algorithm is evaluated
by measuring the VT catalog
completeness and purity as a function of 
mass and redshift. These quantities are 
the selection function
needed to understand the catalog.
The completeness and purity are best measured
with mock galaxy catalogs with known relations 
to dark matter halos. The field can no longer be advanced
by placing single clusters in the center of an 
image with random backgrounds.

We apply the algorithm to a mock galaxy catalog and match the resulting
cluster catalog with the corresponding mock truth table of halos -- the truth
table.  
This allows us to define completeness as the fraction of halos with a 
VT cluster counterpart and purity as the fraction of VT clusters with 
a matching halo. We perform this in bins of redshift and we 
also estimate the impact of redshift errors.

\subsection{Mock catalogs}
Mock galaxy catalogs are created using the ADDGALS code
\citep[see also {\citealt[Appendix A]{Gerdes:2010}}]{Busha:2008,Wechsler:2004}.
ADDGALS takes a N-body simulation light cone and attaches
galaxies to its dark matter particles to create a deep mock photometric catalog using an N-body simulation with only modest mass resolution. The resulting galaxy
catalog reproduces the luminosity function, the magnitude dependent
2-point correlation function and the
color-density-luminosity distribution
measured from the SDSS data.  The mock catalogs used here were based on the Hubble Volume simulation that modeled a 3-Gpc/h box with 1024$^3$ particles in a 
flat $\Lambda$CDM cosmology with $\Omega_M = 0.3$ and $\sigma_8 = 0.9$ \citep{Evrard:2002}.

ADDGALS first builds a list of galaxies $r$-band luminosities
drawing from a luminosity function $\phi(M_r)$, and assigns these galaxies to individual dark matter particles in the simulation. Here, $\phi(M_r)$ is the observed SDSS $r$-band luminosity
function at redshift $\sim 0.1$ from \cite{Blanton:2003} assuming passive evolution of
1.3 magnitudes per unit redshift.
These galaxies are then mapped to individual dark matter particles using a probability relation $P(R_{\delta}|L_r/L_*)$ that relates to local dark matter overdensity to the luminosity of a galaxy.
Overdensities of dark matter are
computed using the characteristic radius $R_{\delta}$, defined  as
the radius enclosing $1.8 \times 10^{13} h^{-1}$ solar masses of
dark matter.
The form of $P(R_{\delta}|L_r/L_*)$ is taken to be a Gaussian plus a log-normal representing galaxies in the ``field'', i.e., unresolved low-mass halos, and those in higher mass, well-resolved ``halos.''  The exact form of this relation 
is
\begin{eqnarray}
P(R_{\delta}l|L_r/L_*) =
{(1-p(L)) \over R\sqrt{2\pi}\sigma_c(L_r/L_*)} \times \nonumber\\
e^{-(\ln(R_{\delta}l)-\mu_c(L_r/L_*))^2/2\sigma_c(L_r/L_*)^2} + \nonumber\\
{p(L_r/L_*) \over \sqrt{2\pi}\sigma_f(:_r/L_*)}
e^{(R_{\delta}-\mu_f(:_r/L_*))^2/2\sigma_f(L_r/L_*)^2}.
\end{eqnarray}
The exact values of the parameters for this function are
determined using a Monte Carlo Markov Chain
analysis, imposing that the observed magnitude
dependent 2-point correlation function is matched.

The next step is to assign galaxy colors. The local galaxy
density is computed for each galaxy in the simulation and in a training set of galaxies from the magnitude-limited SDSS DR6 catalog using the projected 
distance to the 5th nearest neighbor in a bin of redshift as in \cite{Cooper:2007}. Each mock
galaxy is assigned the SED of a randomly selected SDSS galaxy
with similar local galaxy density and absolute magnitude $M_r$.  When doing this matching, we don't match absolute measurements of the densities, but instead 
opt for a relative matching where the SEDs from the densest galaxies in our training set are matched to the densest galaxies in the mock.  This lets up more 
robustly assign SEDs to higher redshift objects where our training set is incomplete.
The SED is then k-corrected and the appropriate filters are applied to
obtain SDSS colors.
At high redshift, color information is extrapolated from low
redshifts: $r$-band magnitudes are passively evolved before selecting the SED from our training-set galaxy which is then $k$-corrected
assuming that the rest-frame colors and the color-density-luminosity
distribution remain unchanged. 

The resulting catalog reproduces
the overall photometric and clustering properties of the
SDSS galaxies at low redshifts ($z \sim 0.3$)
and extends, using simplified assumptions, to
higher redshifts ($z \sim 1.3$) and deeper magnitudes ($r \sim 24$).
The brightest cluster galaxies (BCGs), however are an exception.
BCGs luminosities are tightly correlated with their host halo mass
and are not reproduced by this method.
Therefore, a BCG luminosity is calculated for each resolved halo
(of mass $\sim 5 \times 10^{13}h^{-1}M_{\odot}$ and above) using the measurements from \cite{Hansen:2005}
before the usual galaxy-to-dark matter particle assignment begins.
The corresponding galaxies are then removed from the
initial list of galaxies and placed at the center of its host halo.

We run our cluster finder on  the 
mock catalog and compare our
results with the truth table.
 The quantities
featured in the truth table are R.A., Dec., redshift and
$M_{200}$, plus
list of member galaxies of each halo. In this paper,
we refer to the truth table
as the {\it halo catalog}, and to the VT output as the
{\it cluster catalog}.
The quantities we use as inputs
are: R.A., Dec., and photometric redshift.
We generate photometric redshifts from the true redshifts,
using a Gaussian distribution of width $\sigma_z  (1+z)$. We
test four different values of $\sigma_z$, namely 0.015, 0.03, 0.045 and
0.06 to access the impact of the photometric redshift errors in our
cluster finder.

The discussion so far was restricted to
a perfect volume limited galaxy catalog.
A real galaxy catalog,
however, will have an irreducible level of contamination and
incompleteness. Here we
mimic the effects of
these two quantities in the mocks by
assuming that the input galaxy catalog has a completeness function given by
a Fermi-Dirac distribution
\begin{equation}\label{comp}
C_g(r) = \frac{f_{0}}{1+\exp((r-{\mu})/{\sigma})}
\end{equation}
where $\mu$ is the magnitude limit of the catalog, $f_0$ is a
normalization constant and the parameter $\sigma$ controls how
fast the completeness falls when the magnitude limit is reached.
The parameters $f_0$ and $\sigma$ are taken from processing of the SDSS data
with the 2DPHOT package \citep{LaBarbera:2008}.
We found that $f_0=0.99$ and $\sigma=0.2$ are
typical values. We
degrade the mock catalogs using
$\mu=23.5$, interpreting $C_g(r)$ as
the probability that a galaxy of magnitude $r$ is detected.
Similarly, from the SDSS data we infer that a small fraction of
contaminants, due to misclassified stars, can be present in the input
catalog. The fraction of misclassified objects increases exponentially
for magnitudes above $\mu - 1.5$. We take this fact into account by
generating false galaxies randomly above this limit and drawing
from (\ref{comp}) the probability that this object is actually added
to the catalog.

\subsection{Membership matching}
The evaluation of completeness and purity requires a 
well defined matching scheme between the cluster catalog and 
the truth table. We use a membership-based matching method.
Membership matching 
has been used in evaluating completeness and purity of 
both photometric and spectroscopic catalogs 
\citep{White:2002,Eke:2004,Gerke:2005,Koester:2007}.
Unlike  cylindrical matching, which has been largely employed in 
this kind of study, 
this method is parameter-free, unambiguous and
provides the means to evaluate the efficiency of the cluster
finder as a function of halo mass regardless of the observable 
proxy for mass. This allow us to distinguish the aspects relevant to the 
cluster finding problem from aspects connected to the mass-observable 
proxy calibration, which is a problem per se and is better addressed
by a separate set of post-finding algorithms.

The inputs for the matching code are the halo catalog and the 
cluster catalog. The first is ranked by mass while the latter is ranked by 
the number of galaxies, both in descending order and in bins of redshift. 
It is critical to do the ranking 
in  bins of redshift for both the halos and the clusters.  
In the case of halos, the mass function is evolving, so the masses will be 
changing at fixed rank.  In the case of the clusters, the flux 
limit forces a changing luminosity limit 
with redshift, so the ranks will be changing at fixed mass.  
If this is not taken into account, a massive cluster at high-z ($z\sim1$) 
will get a much lower rank than a massive cluster at low-z ($z\sim0.1$). 
 
After ranking, the first step is to fit a rank-mass relation $R(M)$
to the cluster catalog, provided rank, and 
the matched halo catalog provided mass. We use the fitting formula
\begin{equation}
R(M)=
\left(\frac{M}{M_p}\right)^{\alpha} 
\exp \left(\exp \left( \frac{M_{0} - M}{M_e} \right)
          -\frac{M}{M_1} 
    \right)
\end{equation}
This relation has no motivation other than a global fitting function,
valid at all redshifts  provided that the ranking is performed
as described above. 
For our mock catalogs, the best fit parameters  for this fitting 
function are
$M_p=2.26 \times 10^{17}$,
$M_e=1.40 \times 10^{14}$,
$M_0=1.85 \times 10^{13}$,
$M_1=1.85 \times 10^{14}$ and
$\alpha=-1.15$.
We then invert the relation above to compute an ``observed mass'' 
for each cluster and proceed to the matching.
If the proxy used to rank the clusters has a tight correlation
with mass, the ranking will be accurate and the observed mass will 
show a tight correlation with the true mass for the matched pairs. 
It is important to notice, that the use of  ranking instead of observed mass, 
does not require  the mass-observable relation to be calibrated.
Moreover, neither  mass information nor the ranking 
is used in the matching process, which is membership-based.

A match takes place if a fraction
of member galaxies is shared by a halo-cluster pair. 
The best match is the object sharing the largest fraction of galaxies.
We require unique matching, in which a given halo/cluster is not allowed 
to be associated with more than one cluster/halo.
As both lists are ranked by number of galaxies, 
uniqueness is imposed by eliminating a matched object from the 
list of available objects for future matches down the list.
We also require two-way matching, where the best matching pair is 
found when the matching is performed in both directions, halos-to-clusters
and clusters-to-halos. 


We note that this approach to cluster-halo matching is quite 
general and can be applied to any cluster-finding algorithm that 
produces a list of cluster members. It will be developed in more 
detail as a framework for comparing different algorithms establishing
their usefulness for cosmological tests \citep{Gerke:2011}. 

\subsection{Completeness \& Purity}\label{subsec_comp_pur}
Completeness is defined as the fraction of halos having a counterpart
in the cluster catalog. Purity in turn is defined as the fraction of 
objects in the cluster catalog that correspond to a true halo.
In both cases, only unique two-way matches are considered. 
Allowing for non-unique matching, where each cluster may have more 
than one matching halo and vice-versa, would be a more permissive 
approach. For instance, purity would not be affected 
by a halo being split in two components and completeness would not  
be affected by two halos appearing as a single cluster.  

We count the number of matched objects  in bins of 
mass and redshift. Therefore,
\begin{eqnarray}
C(M,z) = \frac{N_{\mathrm{matched}}(M,z)}{N_{\mathrm{halos}}(M,z)}    \\
P(M,z) =  \frac{N_{\mathrm{matched}}(M,z)}{N_{\mathrm{clusters}}(M,z)}
\end{eqnarray}
Note that $C(M,z)$ can be computed using the true mass of the halos,
being totally independent of the mass proxy used to rank the clusters. 
The true mass of the clusters, however, is available only for the matched 
objects. Therefore $P(M,z)$ has to be computed using the observed
mass and does depend on the ranking. 
We fit a power law to  the 
$M_{\mathrm{obs}}-M_{\mathrm{true}}$ relation from the matched objects
and use it to transform the scale in the $P(M,z)$ plots and show 
both completeness and purity as a function of $M_{\mathrm{true}}$.
This  cannot be performed before the rank-mass relation 
fitting step, which is part of the matching process. 
This method 
allow us to evaluate the efficiency of any
cluster finder imposing minimum requirements, namely  a list of members 
for each cluster.
The selection function can be defined in terms of completeness and 
purity as
\begin{equation}
f(M,z)= \frac{C(M,z)}{P(M,z)}.
\end{equation}
This is a simplified definition.
For cosmological studies with real data, $f(M,z)$ 
should be defined and evaluated in a likelihood analysis that includes the 
scatter in the mass-observable relation after calibration. 
Here, however, we simply want to 
compare the observed cluster  number counts $N_{obs}(M,z)$ to the 
predictions from the $\Lambda$CDM cosmological model 
$N_{\Lambda{\mathrm{CDM}}}(M,z)$. In this case, the selection
function is easily taken into account:
\begin{equation}
N_{obs}(M,z)=f(M,z) N_{\Lambda{\mathrm{CDM}}}(M,z).
\end{equation}
This comparison allows us to 
develop a feel for how 
well we can 
recover the true cluster number counts using the VT catalog and 
 our ability to perform a 
cosmological test  using VT clusters as a probe.

The method described above is very simplified with respect to 
the procedures involved in an actual 
measurement of the mass function. This would require a measurement
of the mass-observable relation and its scatter.
We do not perform this because the VT cluster catalog 
provides only Ngals, the number of galaxies on the membership 
list, as a mass proxy. This Ngals was not optimized to have
a tight relation with mass, as for example the $\lambda$
estimator of \citet{Rozo:2009b}.
Measuring and optimizing a mass proxy 
is a necessary step if the VT is to be used in performing 
cosmological tests. But this problem is better addressed by a
separate algorithm, specifically designed to provide a 
calibrated mass proxy including the mean relation and the scatter. 

\section{Results and discussion}\label{results}
\begin{figure*}[ht]
\begin{center}
\includegraphics[width=120mm]{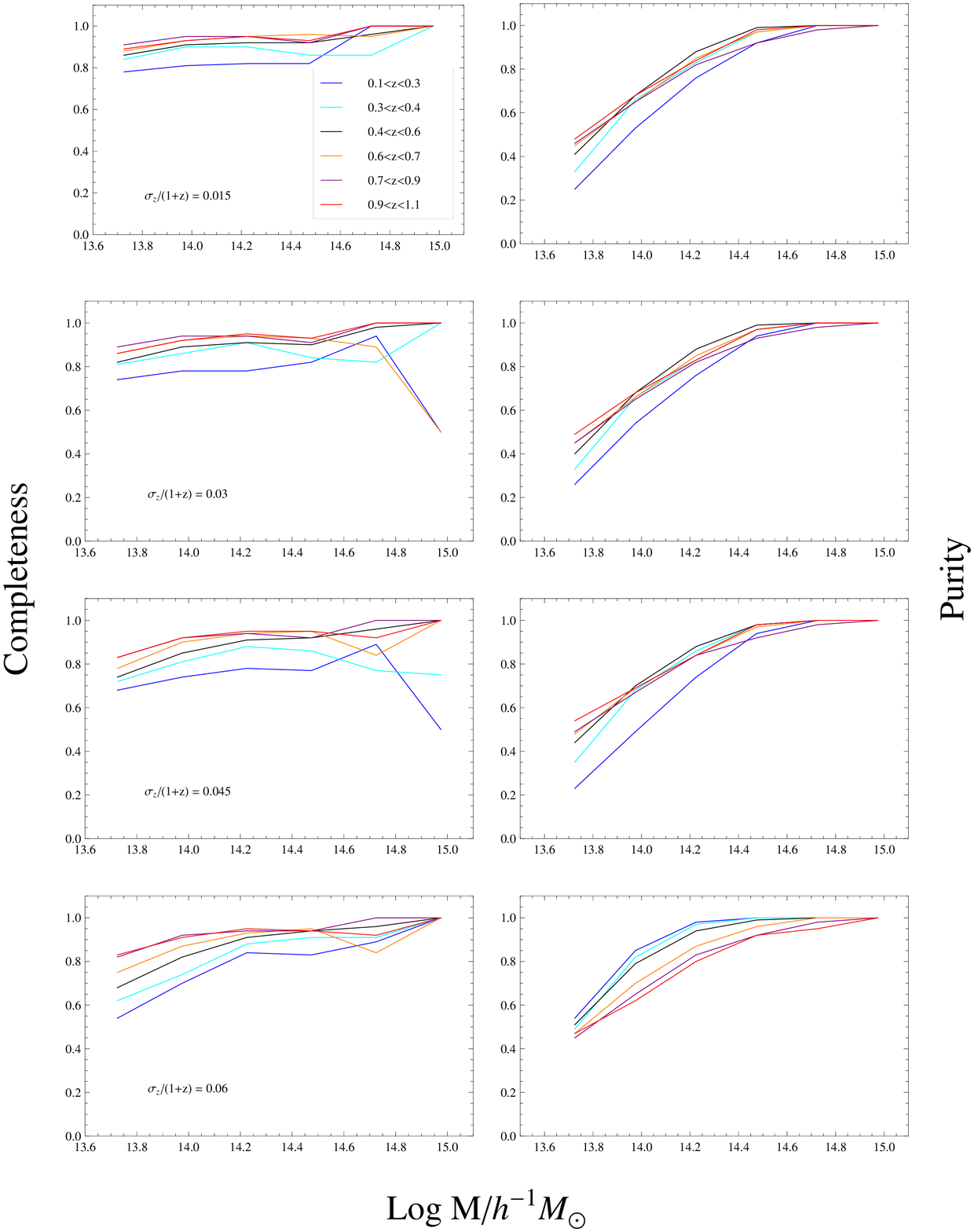}
\end{center}
\caption{Completeness (left) and purity (right) curves as a function of
mass for six redshift bins: 
$0.1<z<0.3$ (blue),   $0.3<z<0.4$ (cyan),   $0.4<z<0.6$ (black),
$0.6<z<0.7$ (orange), $0.7<z<0.9$ (purple), $0.9<z<1.1$ (red).
From top to bottom, the plot pairs feature different $\sigma_z$
values: 0.015, 0.03, 0.045, 0.06. 
The photometric redshift errors have a strong 
impact on both completeness and purity. In the best case, completeness 
and purity rest above 80\% for all redshift bins and 
masses above $\sim 10^{14.2}$. In the case of purity, this curve 
should be interpreted as a lower limit (see text for discussion). 
}
\label{comp_pur1}
\end{figure*}

In Fig.~\ref{comp_pur1} we show the completeness and purity as a function of 
mass and redshift for different Gaussian $\sigma_z$ values.
The photometric redshift errors have a strong 
impact on both completeness and purity. For $\sigma_z=0.015$,
completeness lies above 80\% for all redshift bins and 
masses above $\sim 10^{13.5}M_{\odot}$. 
Purity however, drops significantly at the
low mass end. We attribute this to  
the fact that the range $10^{13.5}-10^{14}M_{\odot}$ 
is at the lower boundary of the 
halo catalogs associated with the mock catalog. 
ADDGALS 
will populate some fraction of real dark 
matter clumps in the simulation even if they are below the 
threshold for detection in the halo catalog.
A fraction of these halos were 
populated with galaxies by ADDGALS, but were not listed in 
the truth table. We have no means to determine the exact fraction 
at this point and therefore we interpret the purity curve as a lower
limit. 

In the high-redshift regime, completeness and purity do not
change much with $\sigma_z$. 
The lowest redshift bin, however, 
shows the lowest purity and completeness in almost
all cases. 
This might be due to the large angular size of clusters
at low-z, as at $z\sim0.1$ the target area of 1 square degree 
corresponds to only a few times the typical $R_{200}$.
However, even in this case the VT catalog achieves
completeness and purity above $\sim80\%$ at all masses.
Since
we are most interested in 
a reliable catalog at high redshifts, we consider the cluster finder
efficiency, as shown in Fig. \ref{comp_pur1}, very good. 

Note that the behavior of purity is qualitatively different in the last 
panel, $\sigma_z=0.060$. This may be connected to low redshift clusters 
leaking to high redshift shells at higher rates than the high redshift 
ones fall towards low redshift.

Testing the effect of changes in the cluster finder free 
parameters  on the completeness and purity functions, we find that:
\begin{enumerate}
\item Changing the fraction of shared galaxies required to consider 
two candidates as the same cluster in the range $40-60$ percent 
has less than 1\% impact on the results. We fix this value at 50\%.
\item The selection function is very sensitive to $scl(z)$.
Setting $scl(z)$ too high ($>0.97$) leads to fragmentation of clusters, which
affects purity at all masses, and failure to detect low contrast 
clusters, which affects completeness at the low mass end.
Setting $scl(z)$ below $0.75$ causes merging of clusters and 
affects completeness. 
An optimal value for $scl(z)$ in the range $0.75-0.97$ 
has to be found at each redshift bin.  
\item The confidence level threshold
has little effect on the detection. The final list of clusters shows  
less than 10\% difference when this parameter varies in the range 
$90-99.5$ percent. But it affects the selection function 
by modifying the membership list.  
\end{enumerate}

\begin{figure}[ht]
\begin{center}
\includegraphics[width=75mm]{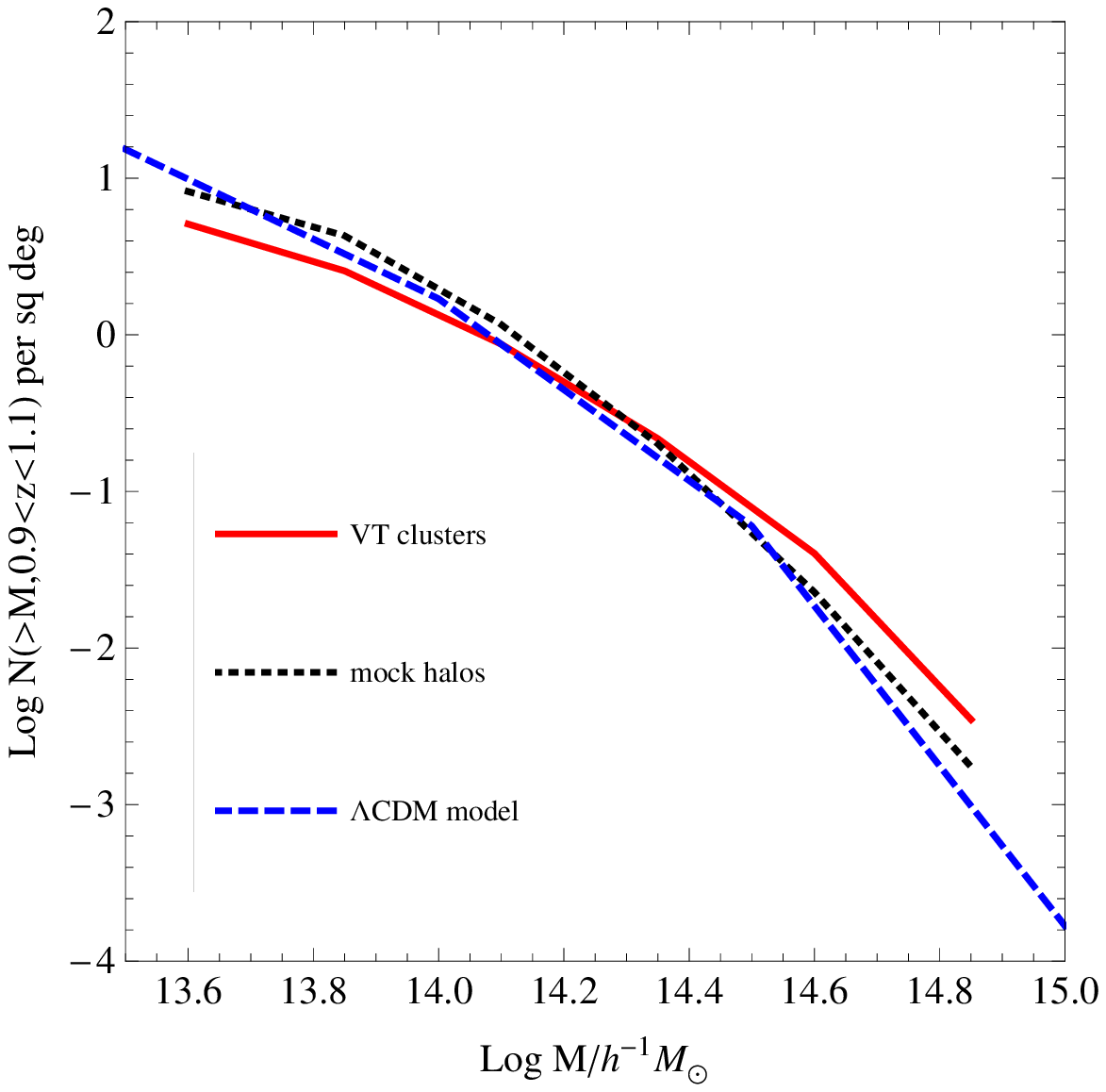}
\end{center}
\caption{Cumulative cluster abundance as a function of mass in the
redshift range $0.9<z<1.1$. The black (dotted) line shows the 
counts in the truth table; the red solid (solid) line shows the 
results of the VT catalog, taking $\sigma_z=0.015$; the blue (dashed)
line shows the values predicted for a $\Lambda$CDM cosmology.
}
\label{mf}
\end{figure}

Fig.~\ref{mf} illustrates our ability to recover the 
true cluster number counts of the input catalog. We take the 
case $\sigma_z=0.015(1+z)$ and the redshift bin $0.9<z<1.1$. For a given 
mass bin $M_i$ we divide the number of VT clusters detected by the selection 
function term $f(M_i,z)$. We then sum the corrected counts through 
all bins of mass $>M$ (red solid line). The curve for the truth table 
is done by counting all the halos above $M$ (black dotted line). 
We finally plot (blue dashed line) the 
values expected in a $\Lambda$CDM cosmology 
\citep[e.g.,][]{Evrard:2002} for comparison.

There is a remarkable agreement between the three curves. 
The tilt of the measured curve with respect to the 
truth table may be interpreted as low mass clusters being misplaced 
towards more massive bins, due to our neglect of the scatter in the 
mass-observable relation.
As pointed out in \S\ref{subsec_comp_pur}, the method used here does
not take into account crucial steps  involved in an actual 
measurement of the mass function. 
This issue must be addressed with a 
full program of mass calibration and is beyond the scope of this 
paper. The result shown in Fig. \ref{mf} encourages the pursuit 
of such a program, though.
Our results  show that the VT is 
a reliable cluster finder in the redshift and mass range of interest,
as seen in the completeness and purity curves.
Application of this algorithm on SDSS data is underway and will be 
presented in a forthcoming paper \citep{Soares-Santos:2010b}.  


\section{Summary}\label{summary}
In this paper we present an improved implementation of the Voronoi
Tessellation cluster finder. Improvements with respect to 
earlier works include:
\begin{enumerate}
\item The use of photometric redshifts instead of magnitudes. 
\item A more realistic assumption that galaxy fields have two-point 
correlation function described by a power-law, and not  by a Poisson
distribution.
\item Implementation of a membership assignment scheme.
\end{enumerate}

The VT cluster finder in 2+1 dimensions was  
 tailored to fulfill the requirements 
of upcoming cosmological experiments aiming at using clusters as 
probes for dark energy.  
The main challenges towards this goal include
the construction of reliable cluster catalogs up to high redshifts 
($z\sim1$) and down to low mass limits ($\sim 10^{13.5}M_{\odot}$)
and the measurement of the selection function as a function of $M$
and $z$. 
To achieve these goals using the VT we:
\begin{enumerate}
\item Adapted the VT algorithm to use photometric redshift shells and 
take advantage of 
the relation that we have discovered 
between the two-point correlation function of the 
galaxy field and its distribution of VT cell areas.
\item Defined the selection function in term of completeness and
purity, establishing  an objective way to measure these quantities 
using simulated catalogs. 
\item Applied the VT to mock  galaxy 
catalogs and computed the completeness and 
purity of the output cluster catalog with the truth table, showing that 
the VT can produce cluster catalogs with completeness and purity above 
80\%  in the ranges of interest within the $M$-$z$ parameter space. 
\item Computed the cluster abundance from the VT catalog and compared 
it to the halo abundance in the mocks, finding a remarkable agreement
at all mass bins. 
\end{enumerate}
These results allow us to be confident in our ability to perform a 
cosmological test for dark energy using the VT algorithm on a 
data set of sufficient scope. Analysis of the application of the VT to 
the SDSS data is underway and will be presented elsewhere.

\acknowledgments
Marcelle Soares-Santos has received support from the Brazilian 
agency CNPq and from the Fermilab Center for Particle 
Astrophysics for this work. 
RHW and BFG received support from the US Department of Energy
under contract number DE-AC02-76SF00515.
Thanks to Massimo Ramella for making his 
code available at http://www.ts.astro.it/astro/VoroHome/ and to 
Yang Jiao for pointing out 
the simulated annealing method applied in this paper. 
Thanks to Jorge Horvath for careful reading of the manuscript.

\bibliographystyle{aas}
\bibliography{bib}

\appendix

\section{The Voronoi Tessellation cell areas distribution for power-law correlated point processes}

Motivated by what is known about the two-point correlation function of 
galaxies in the Universe, we 
consider a 2-dimensional point field characterized by a two-point 
correlation
function of the form 
\begin{equation}\label{wtheta}
w(\theta) = A \theta^{1-\gamma}
\end{equation} 
where $\theta$ is 
a distance, $A$ is the amplitude of the correlation and $\gamma$
is the slope of the power-law.  $A=0$ represents the Poisson particular case.  
We generate simmulated fields spanning a wide range of the parameter space
($A,\gamma$) around the measured values reported in the literature. These
simulated fields are used to characterize the VT cell areas distribution.

Although aimed at application in our cluster finder algorithm, this study 
allows to investigate the connection between this VT property and
the statistical process of the generator set of points. This topic has 
been extensively discussed 
(see \citet{Okabe:2000} for a review). For the Poisson case, simulations 
have been used to support the so-called Kiang's conjecture that the 
distribution of 
standardized cell sizes (size/mean size) in $n$-dimensional space is given by 
\begin{equation}\label{gamma_dist}
p(x) = \frac{\beta^{\alpha}}{\Gamma(\alpha)} x^{\alpha -1 } \exp^{-\beta x}
\end{equation} 
with $\alpha=\beta=2n$. This has been rigorously shown for $n=1$ and 
studied in simulations up to $n=3$. 
Here we extend this conjecture to
the case where the two-point correlation function of the field is given by a
power-law. We focus on $n=2$. Our results indicate 
that Eq.~\ref{gamma_dist} still holds but the
parameters $\alpha$ and $\beta$ are modified. The relation
$\alpha=0.26+\beta$ is found to be valid within the parameters space explored.

In the following sections we describe the simulations and the 
modeling of the area distribution.
We discuss our results in comparison to the well-studied Poisson case and
provide the relevant quantities in Table 1. 
  
\subsection{Point field simulation}
To generate the simulated fields with two-point correlation function given
by Eq.~\ref{wtheta}, we implement the simulated annealing 
method as proposed by \citet{Rintoul:1997}. This method is generally used
to find the state of minimum ``energy'' of a given system, by sampling 
the different states weighted by the probability of occurrence of that 
state. Here, we take Eq.~(\ref{wtheta}) as our ``reference'' state, and
the state of the ``system'' is denoted as $w_s(\theta)$. We consider 
logarithmic bins in $\theta$, and define the energy of the system as 
\begin{equation}
E=\sum_i (w_s(\theta_i)-w(\theta_i))^2
\end{equation}  
where the sum is over all bins. We use 10 bins in
the interval $0.01<\theta_i<2$.
This definition of energy is convenient because it ensures that $E$ decreases
when the difference between any two bins decreases.
 
The initial state is a Poisson state. To evolve the system towards 
$w(\theta)$, we chose a particle and move it to a random position 
in the field. 
We compute the energy $E'$ of this new configuration and obtain 
$\Delta E=E'-E$. The move is accepted with probability 
\begin{equation}
p(\Delta E)=\left\{ \begin{array}{ll}
1                   & \Delta E \leq 0 \\
\exp (-\Delta E/kT) & \Delta E > 0  
\end{array}\right.
\end{equation}
where $kT$ is the ``temperature'' of the system. This is chosen to 
allow the system to evolve as quickly as possible to the  minimum state,
without getting trapped in local minima. The initial temperature is set to
1.
We attempt to move all the particles sequentially and, after a complete
round over all the $N$ particles of the system, its temperature is 
cooled by a factor of 2. The system converges about 30\% faster with this 
cooling schedule.   

\begin{figure}[ht]
\begin{center}
\includegraphics[width=50mm]{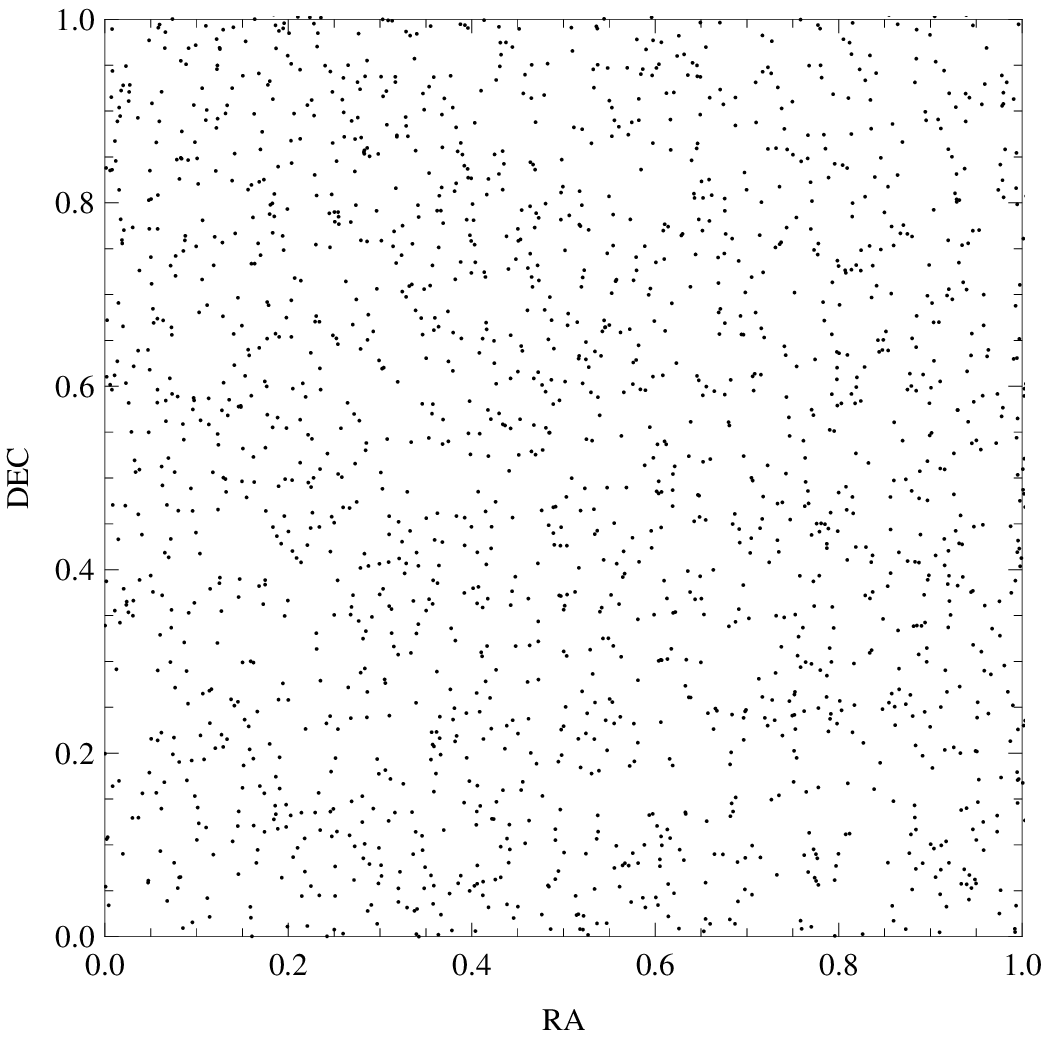}\hspace{5mm}
\includegraphics[width=50mm]{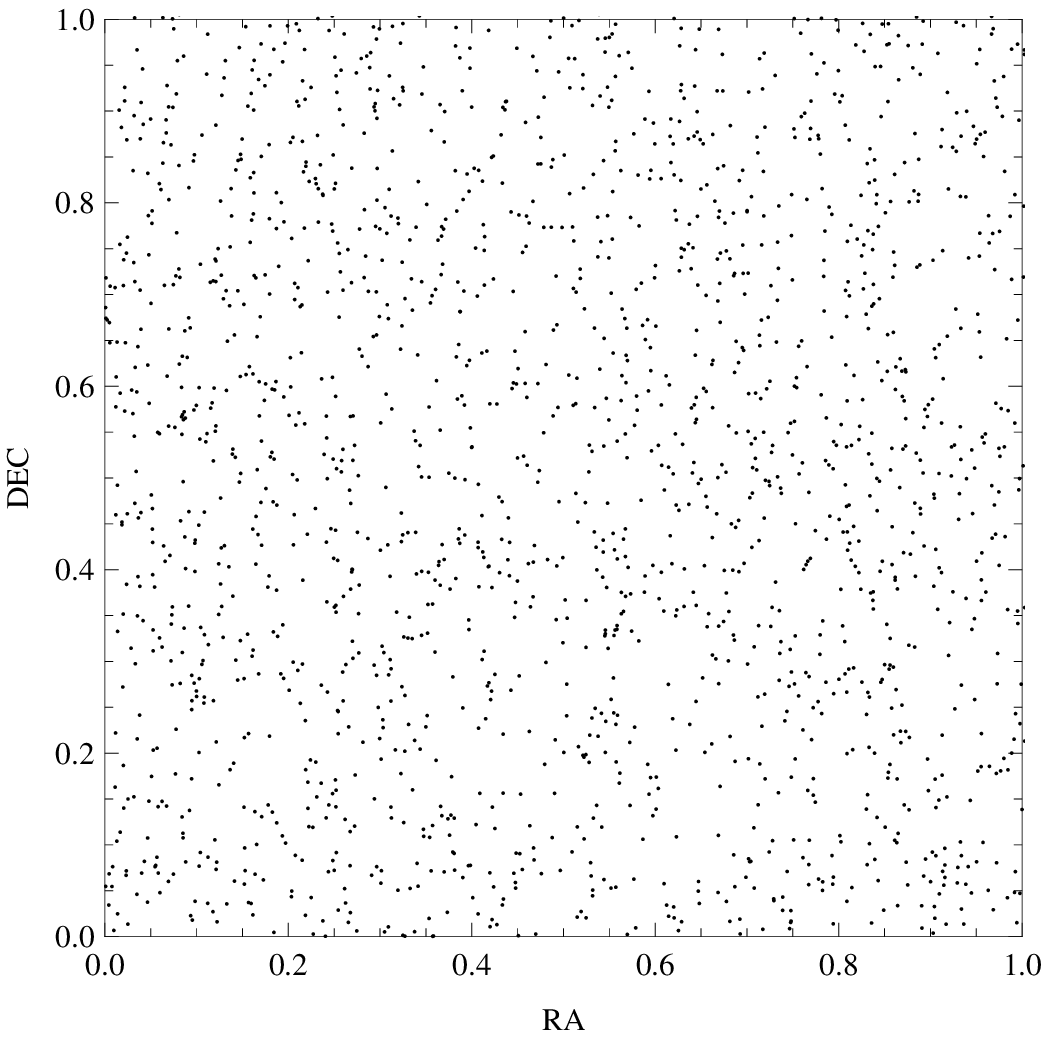}\hspace{5mm}
\includegraphics[width=50mm]{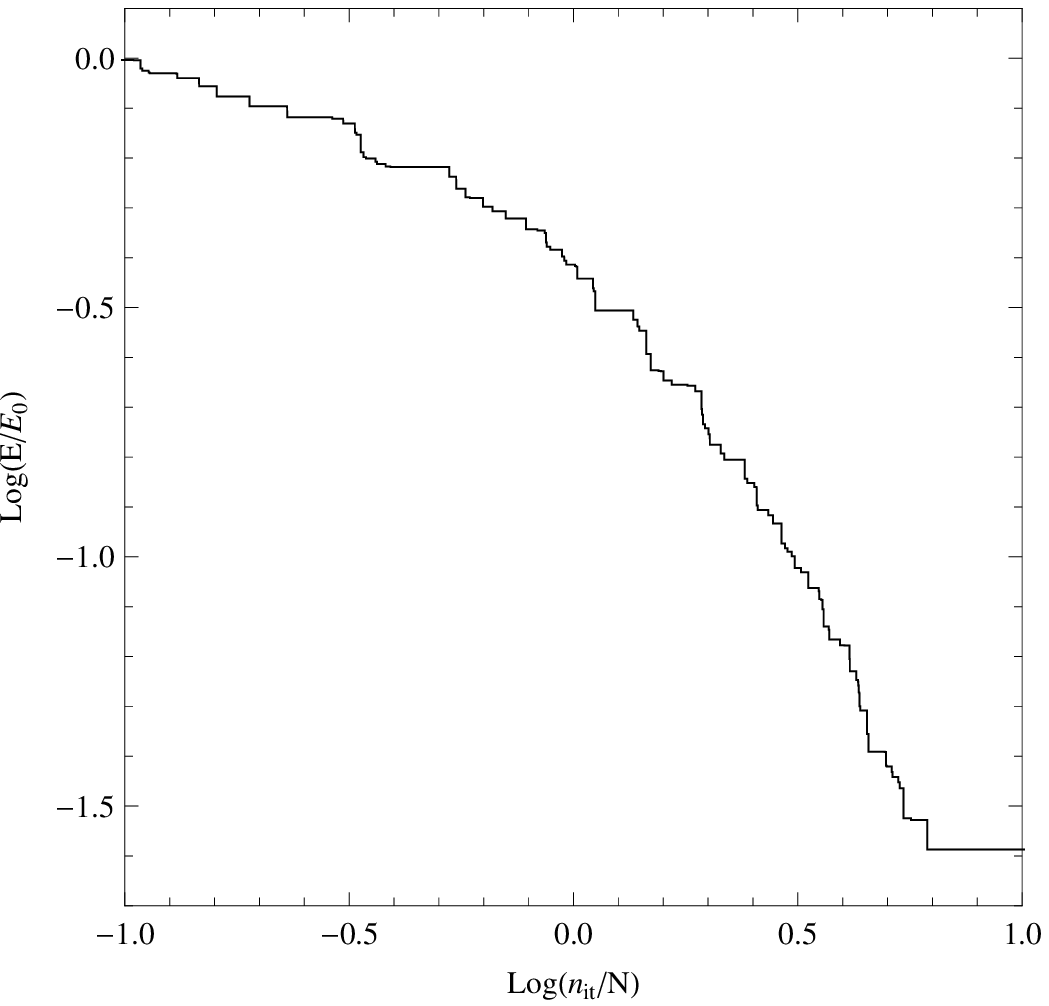}
\end{center}
\caption{The left plot shows the initial (Poisson) state of a system
meant to evolve towards a configuration with $A=0.005$ and $\gamma=1.7$.
The final state is the one in the central plot. The right plot is the
evolution of the energy of the system (normalized by its initial energy)
as a function of the iteration number normalized by the total number of
particles in the system. Under this normalization, $n_{it}/N = 1,2,3...$
refers to complete rounds over all particles in the field. This simulation
was performed in a box of $3\times 3$ sq deg containing $1.6\times 10^4$ 
particles. Just a $1\times 1$ sq deg portion of the field is shown. 
} 
\label{sim_fields}
\end{figure}

In Fig.~\ref{sim_fields} we show one example, where $A=0.005$ and 
$\gamma=1.7$. This combination of parameters correspond to typical values
measured, for instance, on SDSS data up to magnitude limit $r'=21.5$
\citep{Connolly:2002}. 
The initial system is on 
the left, the field in the middle is the final state, after 10 rounds over
all particles. The plot on the right 
shows the evolution of the energy of 
the system. The difference between the initial and final states is not
noticeable by eye and a statistical method must be used to actually
measure the two-point correlation function and compute $\Delta E$ 
at each iteration. We use a fast Fourier transform code 
\citep{Szapudi:2005} to accomplish this. 
Using this method we have generated 190 fields of $3\times3$ sq degrees and
$1.6 \times 10^4$ particles.

\subsection{Gamma model for the VT cell distribution}
\begin{figure}[ht]
\begin{center}
\includegraphics[width=75mm]{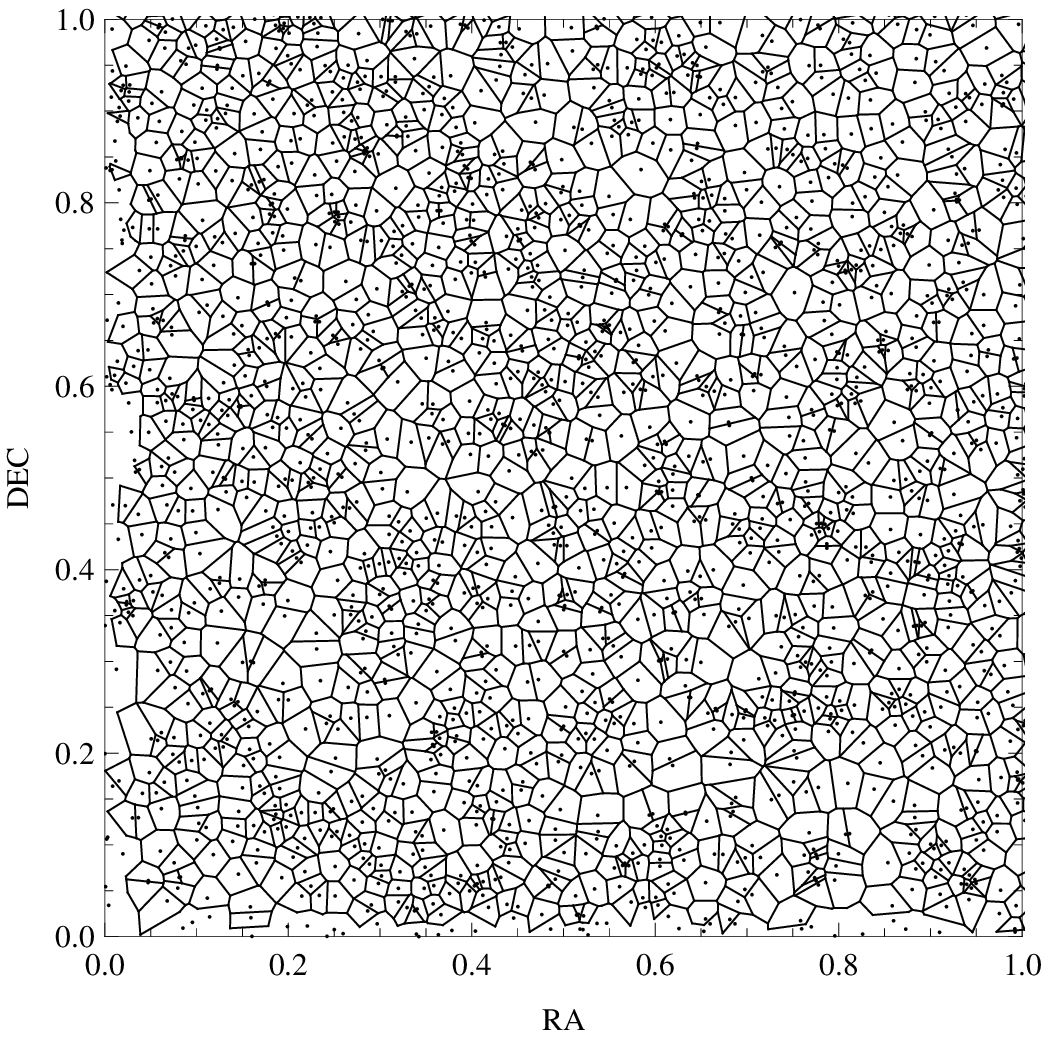}\hspace{5mm}
\includegraphics[width=75mm]{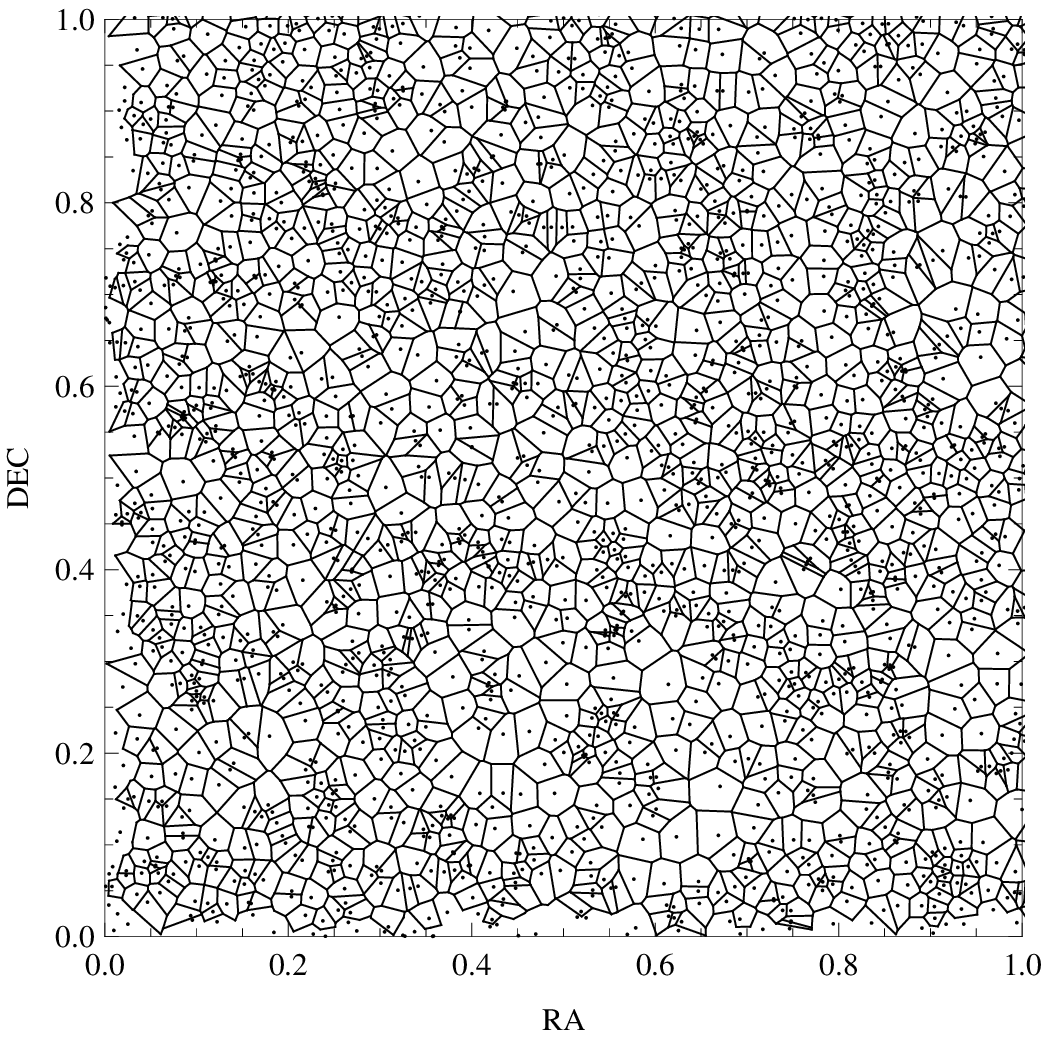}
\end{center}
\caption{The Voronoi diagram corresponding to the two fields shown in 
Fig.~\ref{sim_fields}. The initial and final states are on the left and right
panels, respectively.
}
\label{Poisson_wtheta_VT}
\end{figure}

We apply the VT code on each of the simulated fields, obtain the
distribution of cell normalized cell areas and find the best fit Gamma model
(Eq.~\ref{gamma_dist}). Fig.~\ref{Poisson_wtheta_VT} shows, as an example
the VT diagram for the same system featured above. The left and right diagrams 
correspond to the initial and final state of the system, respectively.  

\begin{figure}[!ht]
\begin{center}
\includegraphics[width=75mm]{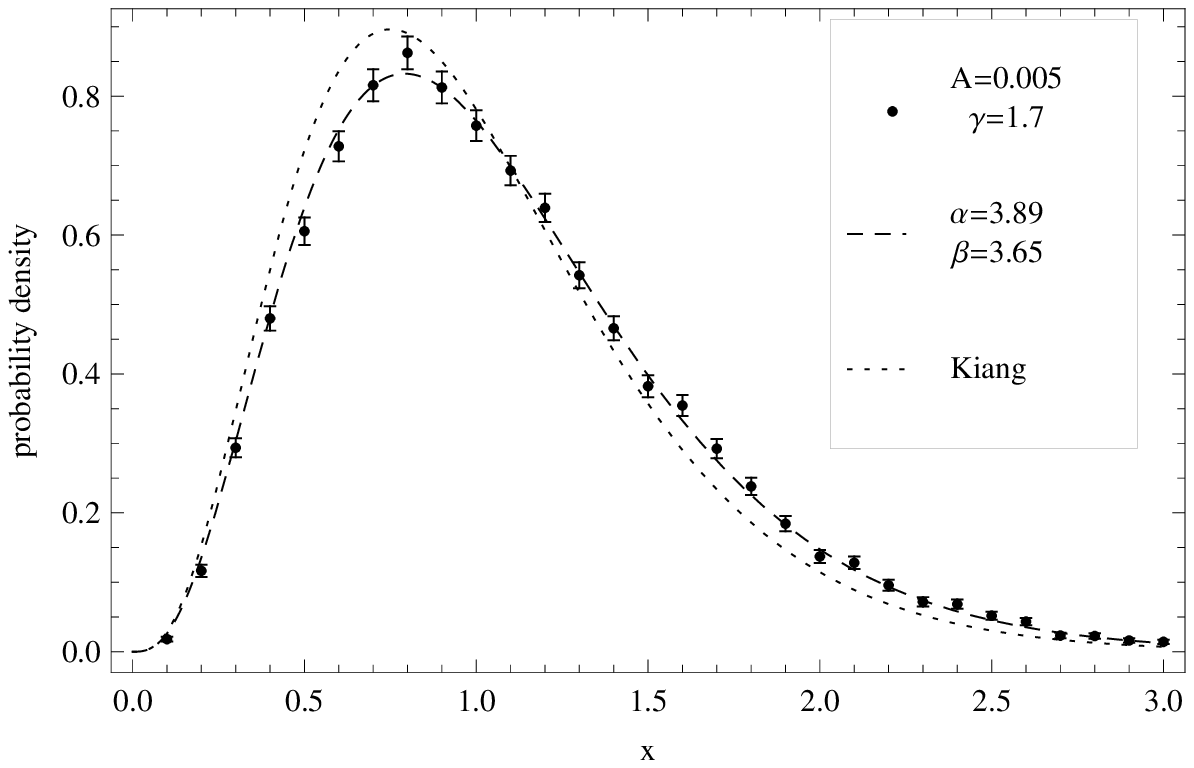}\hspace{5mm}
\includegraphics[width=75mm]{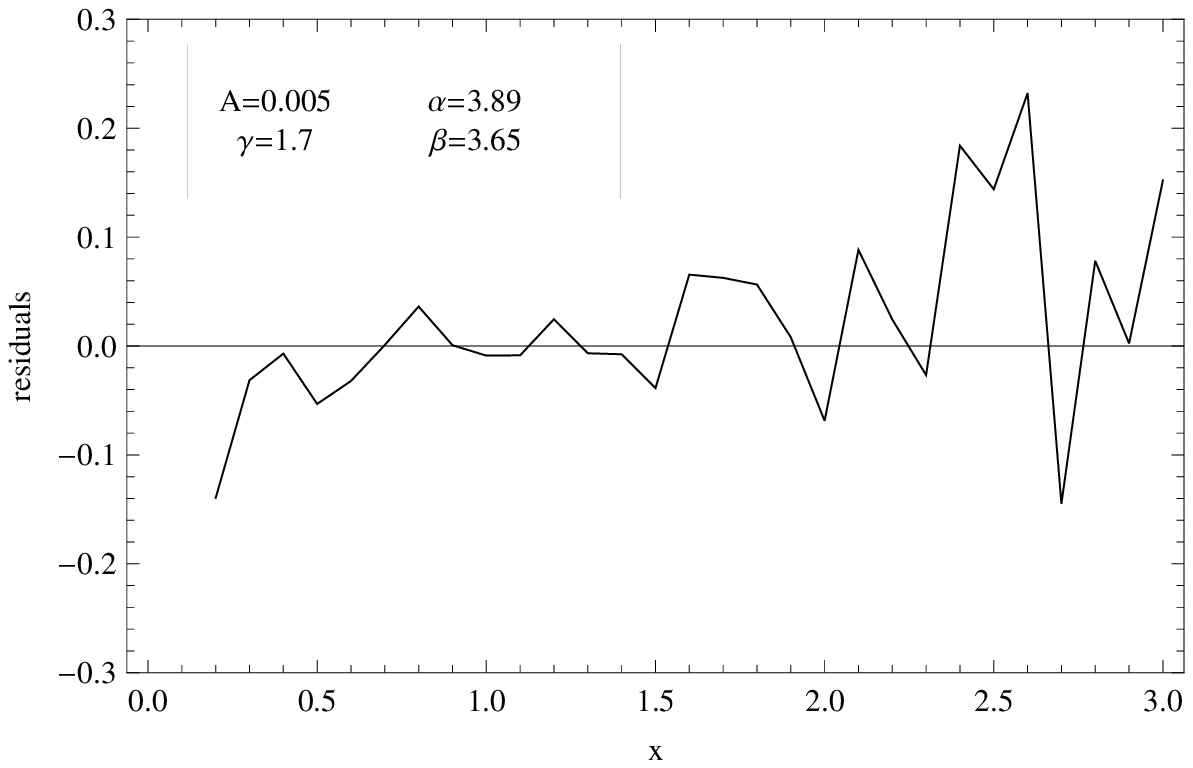}
\end{center}
\caption{Left: Best fit model for the distribution of normalized VT cell areas
featured in Fig.~\ref{Poisson_wtheta_VT}. The curve for the Poisson case 
is also shown for comparison (dashed line). 
Right: Fractional residuals of the fit.
} 
\label{areas_dist}
\end{figure}

The result of the fit is shown in Fig.~\ref{areas_dist}, again for the
case $A=0.005$ and $\gamma=1.7$.  For comparison 
we show as well the traditional Kiang formula (dashed line). The 
results are $\alpha=3.89\pm0.04$ and $\beta=3.65\pm0.05$. Kiang's 
formula is more than 5$\sigma$ away from the best fit. 

\begin{figure}[!ht]
\begin{center}
\includegraphics[width=50mm]{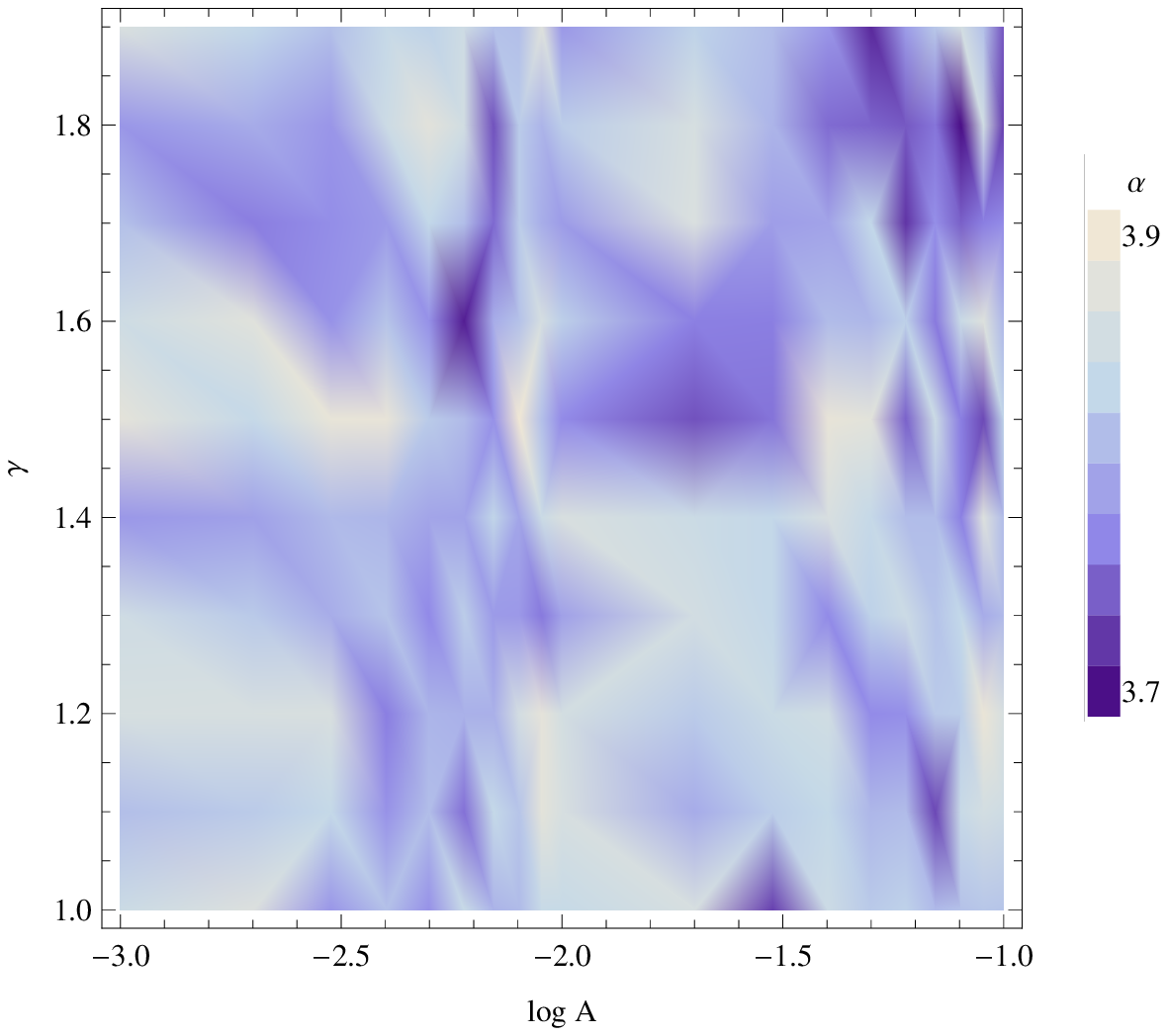}\hspace{1mm}
\includegraphics[width=50mm]{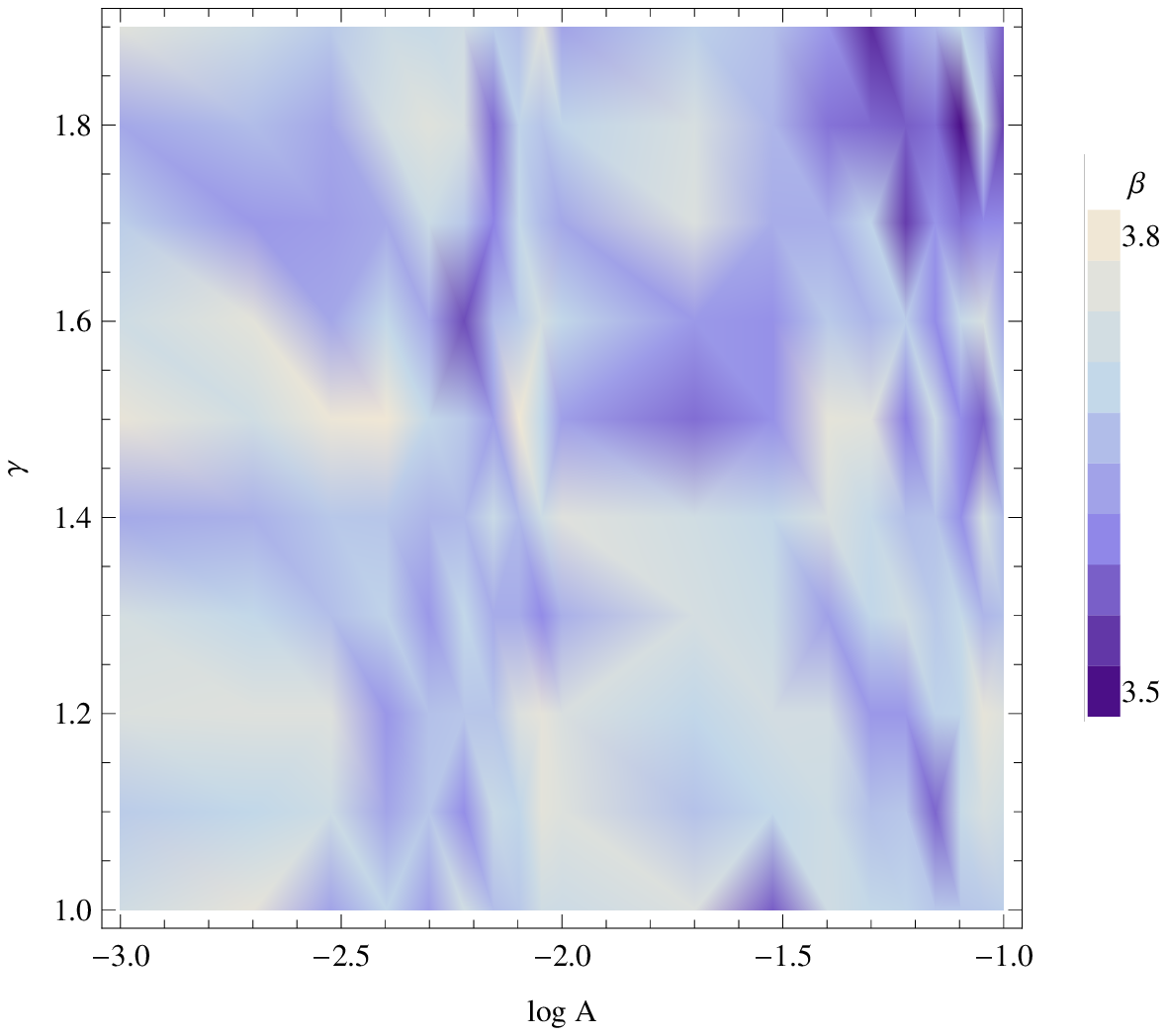}\hspace{1mm}
\includegraphics[width=54mm]{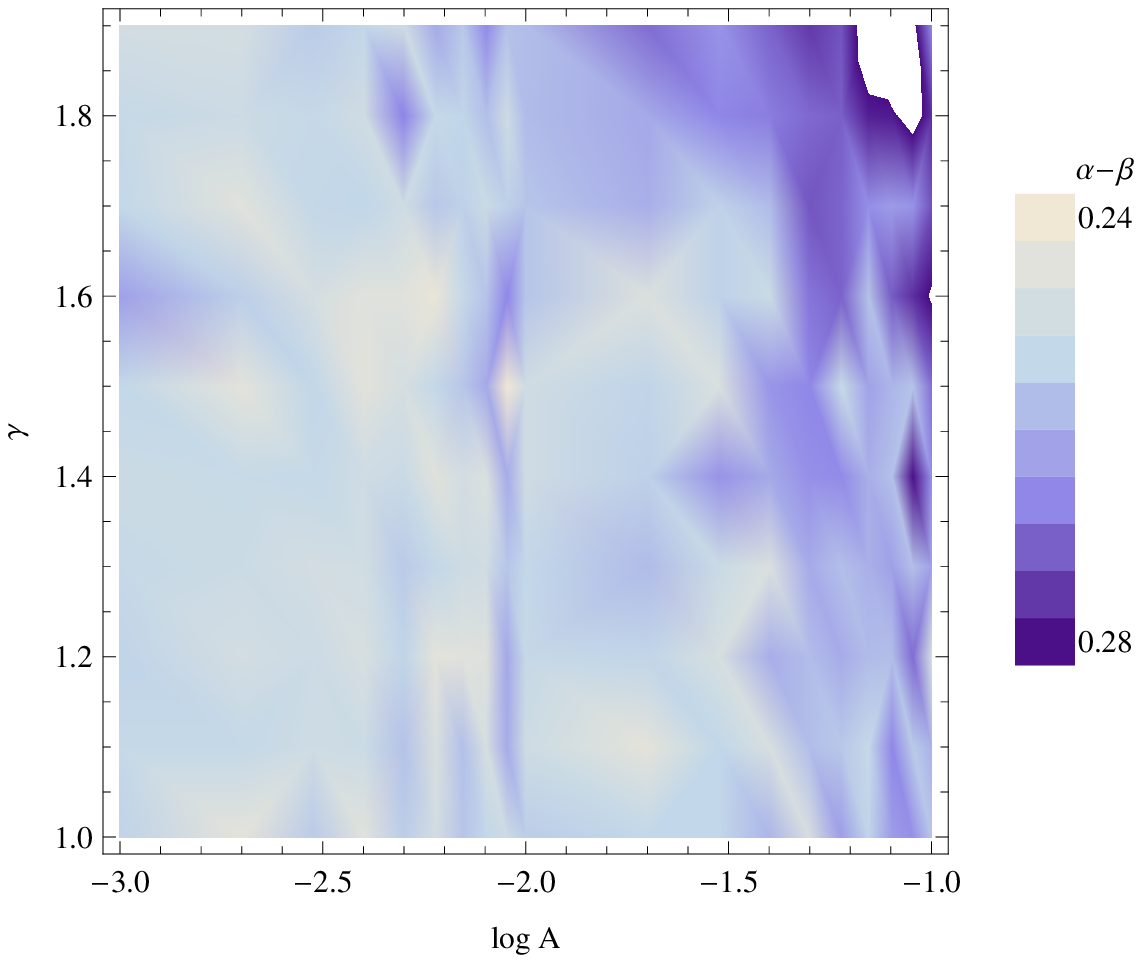}
\end{center}
\caption{Density maps showing the results of the fit in the parameter
space investigated. There is a noticeable correlation between the two
left most maps. The difference  between these two maps is shown in 
the right. 
}
\label{alpha_beta_fitted}
\end{figure}

The results for the ensemble of simulated fields studied are shown in 
Fig.~\ref{alpha_beta_fitted}. The values of $\alpha$ and $\beta$ fall in 
the range $3.5<\alpha<3.9$ and $3.5<\beta<3.8$. The mean error in both 
is 0.04. There is a noticeable correlation between these two parameters.
The difference $\alpha-\beta$ is shown to 
be $0.26\pm0.02$ all over the parameter space explored. 
The model parameters for the values of $A$ and $\gamma$ considered
are presented in Table 1.

\clearpage

\LongTables 
\begin{deluxetable*}{rrrrrrrrrrr} 
\label{VTparamsTable}
\tablecolumns{11}
\tablewidth{0pc}
\tablecaption{VT Model Parameters}
\tablehead{
\multicolumn{2}{c}{}                & \colhead{} & \multicolumn{3}{c}{Areas Distribution}                          \\
\cline{4-6} \cline{8-11} \\
\colhead{$A$} & \colhead{$\gamma$}  & \colhead{} & \colhead{$\alpha$} & \colhead{$\beta$} & \colhead{$\chi^2/\nu$} 
}
\startdata
0.001	&	1.0	&	\nodata	&	3.91	$\pm$	0.05	&	3.66	$\pm$	0.05	&	1.24	  \\
0.001	&	1.1	&	\nodata	&	3.86	$\pm$	0.05	&	3.61	$\pm$	0.05	&	1.49	  \\
0.001	&	1.2	&	\nodata	&	3.93	$\pm$	0.04	&	3.68	$\pm$	0.03	&	1.12	  \\
0.001	&	1.3	&	\nodata	&	3.92	$\pm$	0.04	&	3.67	$\pm$	0.03	&	1.74	  \\
0.001	&	1.4	&	\nodata	&	3.81	$\pm$	0.04	&	3.57	$\pm$	0.04	&	1.7	  \\
0.001	&	1.5	&	\nodata	&	3.96	$\pm$	0.04	&	3.71	$\pm$	0.03	&	1.33	  \\
0.001	&	1.6	&	\nodata	&	3.91	$\pm$	0.04	&	3.65	$\pm$	0.03	&	1.25	  \\
0.001	&	1.7	&	\nodata	&	3.86	$\pm$	0.05	&	3.62	$\pm$	0.05	&	2.07	  \\
0.001	&	1.8	&	\nodata	&	3.81	$\pm$	0.04	&	3.57	$\pm$	0.03	&	1.49	  \\
0.001	&	1.9	&	\nodata	&	3.94	$\pm$	0.02	&	3.69	$\pm$	0.01	&	1.19	  \\
0.002	&	1.0	&	\nodata	&	3.94	$\pm$	0.04	&	3.71	$\pm$	0.03	&	1.78	  \\
0.002	&	1.1	&	\nodata	&	3.87	$\pm$	0.05	&	3.63	$\pm$	0.04	&	1.18	  \\
0.002	&	1.2	&	\nodata	&	3.93	$\pm$	0.04	&	3.69	$\pm$	0.05	&	1.61	  \\
0.002	&	1.3	&	\nodata	&	3.87	$\pm$	0.02	&	3.63	$\pm$	0.03	&	2.17	  \\
0.002	&	1.4	&	\nodata	&	3.83	$\pm$	0.04	&	3.58	$\pm$	0.03	&	1.43	  \\
0.002	&	1.5	&	\nodata	&	3.9	$\pm$	0.04	&	3.66	$\pm$	0.03	&	1.4	 \\
0.002	&	1.6	&	\nodata	&	3.95	$\pm$	0.04	&	3.7	$\pm$	0.03	&	1.36	 \\
0.002	&	1.7	&	\nodata	&	3.79	$\pm$	0.04	&	3.55	$\pm$	0.04	&	1.41	  \\
0.002	&	1.8	&	\nodata	&	3.84	$\pm$	0.04	&	3.59	$\pm$	0.03	&	1.57	  \\
0.002	&	1.9	&	\nodata	&	3.89	$\pm$	0.04	&	3.65	$\pm$	0.05	&	1.57	  \\
0.003	&	1.0	&	\nodata	&	3.81	$\pm$	0.04	&	3.56	$\pm$	0.04	&	1.49	  \\
0.003	&	1.1	&	\nodata	&	3.9	$\pm$	0.04	&	3.65	$\pm$	0.05	&	1.22	  \\
0.003	&	1.2	&	\nodata	&	3.94	$\pm$	0.04	&	3.69	$\pm$	0.03	&	1.51	  \\
0.003	&	1.3	&	\nodata	&	3.84	$\pm$	0.04	&	3.6	$\pm$	0.03	&	1.62	  \\
0.003	&	1.4	&	\nodata	&	3.86	$\pm$	0.01	&	3.61	$\pm$	0.01	&	1.53	  \\
0.003	&	1.5	&	\nodata	&	3.97	$\pm$	0.04	&	3.72	$\pm$	0.03	&	1.29	 \\
0.003	&	1.6	&	\nodata	&	3.81	$\pm$	0.04	&	3.57	$\pm$	0.04	&	1.89	  \\
0.003	&	1.7	&	\nodata	&	3.8	$\pm$	0.04	&	3.55	$\pm$	0.05	&	2.03	  \\
0.003	&	1.8	&	\nodata	&	3.81	$\pm$	0.05	&	3.57	$\pm$	0.04	&	1.49	  \\
0.003	&	1.9	&	\nodata	&	3.86	$\pm$	0.04	&	3.61	$\pm$	0.05	&	1.56	  \\
0.004	&	1.0	&	\nodata	&	3.86	$\pm$	0.04	&	3.62	$\pm$	0.05	&	1.59	  \\
0.004	&	1.1	&	\nodata	&	3.81	$\pm$	0.04	&	3.56	$\pm$	0.05	&	1.47	  \\
0.004	&	1.2	&	\nodata	&	3.79	$\pm$	0.04	&	3.55	$\pm$	0.04	&	1.35	  \\
0.004	&	1.3	&	\nodata	&	3.87	$\pm$	0.04	&	3.62	$\pm$	0.03	&	1.65	  \\
0.004	&	1.4	&	\nodata	&	3.85	$\pm$	0.04	&	3.6	$\pm$	0.05	&	1.42	  \\
0.004	&	1.5	&	\nodata	&	3.97	$\pm$	0.04	&	3.73	$\pm$	0.03	&	1.24	  \\
0.004	&	1.6	&	\nodata	&	3.87	$\pm$	0.05	&	3.63	$\pm$	0.05	&	1.35	  \\
0.004	&	1.7	&	\nodata	&	3.82	$\pm$	0.04	&	3.57	$\pm$	0.04	&	1.38	  \\
0.004	&	1.8	&	\nodata	&	3.91	$\pm$	0.04	&	3.66	$\pm$	0.03	&	1.04	  \\
0.004	&	1.9	&	\nodata	&	3.9	$\pm$	0.02	&	3.65	$\pm$	0.01	&	1.33	  \\
0.005	&	1.0	&	\nodata	&	3.81	$\pm$	0.01	&	3.56	$\pm$	0.03	&	1.51	 \\*
0.005	&	1.1	&	\nodata	&	3.86	$\pm$	0.04	&	3.61	$\pm$	0.03	&	1.53	 \\
0.005	&	1.2	&	\nodata	&	3.85	$\pm$	0.04	&	3.6	$\pm$	0.05	&	1.61	 \\
0.005	&	1.3	&	\nodata	&	3.8	$\pm$	0.04	&	3.55	$\pm$	0.04	&	1.41	  \\
0.005	&	1.4	&	\nodata	&	3.83	$\pm$	0.04	&	3.58	$\pm$	0.03	&	1.71	  \\
0.005	&	1.5	&	\nodata	&	3.87	$\pm$	0.04	&	3.63	$\pm$	0.05	&	1.25	  \\
0.005	&	1.6	&	\nodata	&	3.81	$\pm$	0.04	&	3.57	$\pm$	0.03	&	1.14	  \\
0.005	&	1.7	&	\nodata	&	3.89	$\pm$	0.04	&	3.65	$\pm$	0.05	&	1.16	  \\
0.005	&	1.8	&	\nodata	&	3.96	$\pm$	0.04	&	3.69	$\pm$	0.05	&	1.84	  \\
0.005	&	1.9	&	\nodata	&	3.88	$\pm$	0.04	&	3.64	$\pm$	0.05	&	1.56	  \\
0.006	&	1.0	&	\nodata	&	3.9	$\pm$	0.05	&	3.66	$\pm$	0.04	&	1.56	  \\
0.006	&	1.1	&	\nodata	&	3.78	$\pm$	0.01	&	3.54	$\pm$	0.03	&	1.47	  \\
0.006	&	1.2	&	\nodata	&	3.84	$\pm$	0.04	&	3.61	$\pm$	0.03	&	1.13	  \\
0.006	&	1.3	&	\nodata	&	3.88	$\pm$	0.05	&	3.63	$\pm$	0.05	&	1.48	  \\
0.006	&	1.4	&	\nodata	&	3.83	$\pm$	0.01	&	3.59	$\pm$	0.01	&	2.	 \\
0.006	&	1.5	&	\nodata	&	3.86	$\pm$	0.04	&	3.61	$\pm$	0.05	&	1.62	  \\
0.006	&	1.6	&	\nodata	&	3.71	$\pm$	0.04	&	3.47	$\pm$	0.04	&	2.34	  \\
0.006	&	1.7	&	\nodata	&	3.86	$\pm$	0.02	&	3.61	$\pm$	0.03	&	1.62	  \\
0.006	&	1.8	&	\nodata	&	3.92	$\pm$	0.05	&	3.67	$\pm$	0.05	&	1.34	  \\
0.006	&	1.9	&	\nodata	&	3.91	$\pm$	0.02	&	3.66	$\pm$	0.01	&	1.25	  \\
0.007	&	1.0	&	\nodata	&	3.85	$\pm$	0.05	&	3.6	$\pm$	0.04	&	1.53	  \\
0.007	&	1.1	&	\nodata	&	3.9	$\pm$	0.04	&	3.64	$\pm$	0.05	&	2.08	  \\
0.007	&	1.2	&	\nodata	&	3.84	$\pm$	0.01	&	3.6	$\pm$	0.03	&	1.13	  \\
0.007	&	1.3	&	\nodata	&	3.82	$\pm$	0.04	&	3.57	$\pm$	0.03	&	1.51	  \\
0.007	&	1.4	&	\nodata	&	3.89	$\pm$	0.02	&	3.64	$\pm$	0.001	&	1.43	  \\
0.007	&	1.5	&	\nodata	&	3.81	$\pm$	0.01	&	3.56	$\pm$	0.01	&	2.11	  \\
0.007	&	1.6	&	\nodata	&	3.84	$\pm$	0.05	&	3.59	$\pm$	0.05	&	1.53	  \\
0.007	&	1.7	&	\nodata	&	3.77	$\pm$	0.01	&	3.52	$\pm$	0.001	&	1.29	  \\
0.007	&	1.8	&	\nodata	&	3.75	$\pm$	0.04	&	3.5	$\pm$	0.04	&	1.91	 \\
0.007	&	1.9	&	\nodata	&	3.86	$\pm$	0.05	&	3.61	$\pm$	0.04	&	1.52	 \\
0.008	&	1.0	&	\nodata	&	3.86	$\pm$	0.04	&	3.61	$\pm$	0.05	&	1.89	  \\
0.008	&	1.1	&	\nodata	&	3.87	$\pm$	0.04	&	3.62	$\pm$	0.05	&	1.53	  \\
0.008	&	1.2	&	\nodata	&	3.92	$\pm$	0.05	&	3.68	$\pm$	0.05	&	1.67	  \\
0.008	&	1.3	&	\nodata	&	3.82	$\pm$	0.04	&	3.57	$\pm$	0.03	&	1.32	  \\
0.008	&	1.4	&	\nodata	&	3.83	$\pm$	0.04	&	3.59	$\pm$	0.03	&	1.66	  \\
0.008	&	1.5	&	\nodata	&	3.99	$\pm$	0.04	&	3.73	$\pm$	0.03	&	1.67	  \\
0.008	&	1.6	&	\nodata	&	3.86	$\pm$	0.02	&	3.61	$\pm$	0.03	&	1.46	  \\
0.008	&	1.7	&	\nodata	&	3.88	$\pm$	0.05	&	3.63	$\pm$	0.05	&	1.52	  \\
0.008	&	1.8	&	\nodata	&	3.88	$\pm$	0.05	&	3.62	$\pm$	0.05	&	1.39	  \\
0.008	&	1.9	&	\nodata	&	3.86	$\pm$	0.04	&	3.6	$\pm$	0.03	&	1.34	  \\
0.009	&	1.0	&	\nodata	&	3.9	$\pm$	0.02	&	3.66	$\pm$	0.03	&	1.25	\\
0.009	&	1.1	&	\nodata	&	3.96	$\pm$	0.05	&	3.7	$\pm$	0.05	&	1.48	  \\
0.009	&	1.2	&	\nodata	&	3.96	$\pm$	0.04	&	3.71	$\pm$	0.05	&	1.51	  \\
0.009	&	1.3	&	\nodata	&	3.78	$\pm$	0.05	&	3.53	$\pm$	0.04	&	2.21	  \\
0.009	&	1.4	&	\nodata	&	3.91	$\pm$	0.04	&	3.65	$\pm$	0.05	&	1.47	 \\
0.009	&	1.5	&	\nodata	&	3.86	$\pm$	0.04	&	3.63	$\pm$	0.05	&	1.32	  \\
0.009	&	1.6	&	\nodata	&	3.93	$\pm$	0.02	&	3.67	$\pm$	0.01	&	1.21	  \\
0.009	&	1.7	&	\nodata	&	3.84	$\pm$	0.04	&	3.59	$\pm$	0.04	&	1.3	  \\
0.009	&	1.8	&	\nodata	&	3.85	$\pm$	0.04	&	3.6	$\pm$	0.03	&	1.65	  \\
0.009	&	1.9	&	\nodata	&	3.95	$\pm$	0.05	&	3.69	$\pm$	0.04	&	1.44	 \\
0.01	&	1.0	&	\nodata	&	3.9	$\pm$	0.02	&	3.65	$\pm$	0.03	&	1.71	  \\
0.01	&	1.1	&	\nodata	&	3.93	$\pm$	0.04	&	3.69	$\pm$	0.03	&	1.34	 \\
0.01	&	1.2	&	\nodata	&	3.93	$\pm$	0.04	&	3.68	$\pm$	0.05	&	1.5	  \\
0.01	&	1.3	&	\nodata	&	3.83	$\pm$	0.05	&	3.58	$\pm$	0.04	&	1.74	  \\
0.01	&	1.4	&	\nodata	&	3.94	$\pm$	0.05	&	3.69	$\pm$	0.05	&	1.37	  \\
0.01	&	1.5	&	\nodata	&	3.8	$\pm$	0.04	&	3.56	$\pm$	0.04	&	1.3	  \\
0.01	&	1.6	&	\nodata	&	3.88	$\pm$	0.02	&	3.63	$\pm$	0.03	&	1.26	  \\
0.01	&	1.7	&	\nodata	&	3.82	$\pm$	0.05	&	3.57	$\pm$	0.04	&	1.82	  \\
0.01	&	1.8	&	\nodata	&	3.88	$\pm$	0.02	&	3.62	$\pm$	0.03	&	1.52	  \\
0.01	&	1.9	&	\nodata	&	3.82	$\pm$	0.01	&	3.56	$\pm$	0.03	&	1.49	 \\
0.02	&	1.0	&	\nodata	&	3.93	$\pm$	0.04	&	3.68	$\pm$	0.05	&	1.97	  \\
0.02	&	1.1	&	\nodata	&	3.84	$\pm$	0.01	&	3.6	$\pm$	0.01	&	1.96	  \\
0.02	&	1.2	&	\nodata	&	3.88	$\pm$	0.04	&	3.63	$\pm$	0.03	&	1.2	  \\
0.02	&	1.3	&	\nodata	&	3.92	$\pm$	0.04	&	3.67	$\pm$	0.03	&	1.4	 \\
0.02	&	1.4	&	\nodata	&	3.91	$\pm$	0.05	&	3.66	$\pm$	0.05	&	1.51	\\
0.02	&	1.5	&	\nodata	&	3.75	$\pm$	0.01	&	3.5	$\pm$	0.01	&	1.72	  \\
0.02	&	1.6	&	\nodata	&	3.79	$\pm$	0.04	&	3.55	$\pm$	0.03	&	1.2	  \\
0.02	&	1.7	&	\nodata	&	3.94	$\pm$	0.04	&	3.68	$\pm$	0.05	&	1.35	  \\
0.02	&	1.8	&	\nodata	&	3.93	$\pm$	0.04	&	3.67	$\pm$	0.03	&	1.31	  \\
0.02	&	1.9	&	\nodata	&	3.88	$\pm$	0.04	&	3.62	$\pm$	0.03	&	1.09	 \\
0.03	&	1.0	&	\nodata	&	3.74	$\pm$	0.01	&	3.49	$\pm$	0.03	&	1.85	  \\
0.03	&	1.1	&	\nodata	&	3.88	$\pm$	0.05	&	3.63	$\pm$	0.05	&	1.61	  \\
0.03	&	1.2	&	\nodata	&	3.91	$\pm$	0.02	&	3.66	$\pm$	0.03	&	1.66	  \\
0.03	&	1.3	&	\nodata	&	3.89	$\pm$	0.05	&	3.65	$\pm$	0.04	&	1.91	  \\
0.03	&	1.4	&	\nodata	&	3.89	$\pm$	0.02	&	3.63	$\pm$	0.03	&	1.55	\\
0.03	&	1.5	&	\nodata	&	3.78	$\pm$	0.01	&	3.54	$\pm$	0.03	&	1.22	  \\
0.03	&	1.6	&	\nodata	&	3.79	$\pm$	0.04	&	3.54	$\pm$	0.04	&	1.48	  \\
0.03	&	1.7	&	\nodata	&	3.82	$\pm$	0.04	&	3.57	$\pm$	0.03	&	1.52	  \\
0.03	&	1.8	&	\nodata	&	3.85	$\pm$	0.02	&	3.59	$\pm$	0.01	&	0.998	  \\
0.03	&	1.9	&	\nodata	&	3.86	$\pm$	0.04	&	3.6	$\pm$	0.05	&	1.08	  \\
0.04	&	1.0	&	\nodata	&	3.91	$\pm$	0.05	&	3.66	$\pm$	0.05	&	1.53	 \\
0.04	&	1.1	&	\nodata	&	3.89	$\pm$	0.04	&	3.65	$\pm$	0.05	&	1.6	  \\
0.04	&	1.2	&	\nodata	&	3.92	$\pm$	0.04	&	3.66	$\pm$	0.05	&	1.73	  \\
0.04	&	1.3	&	\nodata	&	3.8	$\pm$	0.01	&	3.56	$\pm$	0.03	&	1.68	  \\
0.04	&	1.4	&	\nodata	&	3.93	$\pm$	0.04	&	3.68	$\pm$	0.03	&	1.03	 \\
0.04	&	1.5	&	\nodata	&	3.97	$\pm$	0.04	&	3.71	$\pm$	0.03	&	1.22	 \\
0.04	&	1.6	&	\nodata	&	3.86	$\pm$	0.02	&	3.61	$\pm$	0.01	&	1.39	  \\
0.04	&	1.7	&	\nodata	&	3.83	$\pm$	0.01	&	3.57	$\pm$	0.03	&	1.23	  \\
0.04	&	1.8	&	\nodata	&	3.77	$\pm$	0.04	&	3.51	$\pm$	0.03	&	1.18	  \\
0.04	&	1.9	&	\nodata	&	3.81	$\pm$	0.04	&	3.54	$\pm$	0.03	&	1.32	  \\
0.05	&	1.0	&	\nodata	&	3.87	$\pm$	0.04	&	3.63	$\pm$	0.05	&	0.971	  \\
0.05	&	1.1	&	\nodata	&	3.85	$\pm$	0.01	&	3.6	$\pm$	0.03	&	1.33	 \\
0.05	&	1.2	&	\nodata	&	3.8	$\pm$	0.04	&	3.55	$\pm$	0.03	&	1.18	  \\
0.05	&	1.3	&	\nodata	&	3.88	$\pm$	0.04	&	3.63	$\pm$	0.03	&	1.39	  \\
0.05	&	1.4	&	\nodata	&	3.9	$\pm$	0.04	&	3.64	$\pm$	0.03	&	1.29	  \\
0.05	&	1.5	&	\nodata	&	3.96	$\pm$	0.04	&	3.69	$\pm$	0.05	&	1.27	  \\
0.05	&	1.6	&	\nodata	&	3.85	$\pm$	0.04	&	3.59	$\pm$	0.05	&	1.42	  \\
0.05	&	1.7	&	\nodata	&	3.89	$\pm$	0.05	&	3.62	$\pm$	0.05	&	1.19	  \\
0.05	&	1.8	&	\nodata	&	3.77	$\pm$	0.04	&	3.5	$\pm$	0.04	&	1.27	  \\
0.05	&	1.9	&	\nodata	&	3.72	$\pm$	0.01	&	3.45	$\pm$	0.01	&	1.45	  \\
0.06	&	1.0	&	\nodata	&	3.88	$\pm$	0.04	&	3.62	$\pm$	0.05	&	1.36	  \\
0.06	&	1.1	&	\nodata	&	3.86	$\pm$	0.04	&	3.61	$\pm$	0.03	&	1.62	  \\
0.06	&	1.2	&	\nodata	&	3.8	$\pm$	0.04	&	3.55	$\pm$	0.03	&	1.37	 \\
0.06	&	1.3	&	\nodata	&	3.91	$\pm$	0.02	&	3.66	$\pm$	0.03	&	1.72	  \\
0.06	&	1.4	&	\nodata	&	3.86	$\pm$	0.04	&	3.6	$\pm$	0.03	&	1.19	  \\
0.06	&	1.5	&	\nodata	&	3.77	$\pm$	0.04	&	3.52	$\pm$	0.04	&	1.6	  \\
0.06	&	1.6	&	\nodata	&	3.88	$\pm$	0.04	&	3.61	$\pm$	0.03	&	1.31	  \\
0.06	&	1.7	&	\nodata	&	3.73	$\pm$	0.04	&	3.46	$\pm$	0.03	&	1.52	 \\
0.06	&	1.8	&	\nodata	&	3.76	$\pm$	0.01	&	3.49	$\pm$	0.03	&	1.14	 \\
0.06	&	1.9	&	\nodata	&	3.82	$\pm$	0.04	&	3.55	$\pm$	0.05	&	1.29	 \\
0.07	&	1.0	&	\nodata	&	3.85	$\pm$	0.04	&	3.61	$\pm$	0.03	&	1.58	  \\
0.07	&	1.1	&	\nodata	&	3.74	$\pm$	0.01	&	3.5	$\pm$	0.01	&	1.28	  \\
0.07	&	1.2	&	\nodata	&	3.87	$\pm$	0.02	&	3.62	$\pm$	0.03	&	1.67	 \\
0.07	&	1.3	&	\nodata	&	3.87	$\pm$	0.04	&	3.61	$\pm$	0.03	&	1.47	\\
0.07	&	1.4	&	\nodata	&	3.86	$\pm$	0.05	&	3.6	$\pm$	0.05	&	1.8	  \\
0.07	&	1.5	&	\nodata	&	3.91	$\pm$	0.05	&	3.65	$\pm$	0.05	&	1.34	  \\
0.07	&	1.6	&	\nodata	&	3.78	$\pm$	0.04	&	3.53	$\pm$	0.03	&	1.36	  \\
0.07	&	1.7	&	\nodata	&	3.8	$\pm$	0.04	&	3.54	$\pm$	0.04	&	1.49	  \\
0.07	&	1.8	&	\nodata	&	3.78	$\pm$	0.01	&	3.5	$\pm$	0.03	&	1.72	  \\
0.07	&	1.9	&	\nodata	&	3.87	$\pm$	0.02	&	3.59	$\pm$	0.01	&	1.01	  \\
0.08	&	1.0	&	\nodata	&	3.86	$\pm$	0.04	&	3.6	$\pm$	0.03	&	1.42	  \\
0.08	&	1.1	&	\nodata	&	3.9	$\pm$	0.05	&	3.64	$\pm$	0.04	&	1.28	  \\
0.08	&	1.2	&	\nodata	&	3.88	$\pm$	0.04	&	3.62	$\pm$	0.03	&	1.03	  \\
0.08	&	1.3	&	\nodata	&	3.89	$\pm$	0.05	&	3.63	$\pm$	0.05	&	1.47	 \\
0.08	&	1.4	&	\nodata	&	3.79	$\pm$	0.04	&	3.54	$\pm$	0.04	&	1.53	  \\
0.08	&	1.5	&	\nodata	&	3.79	$\pm$	0.05	&	3.53	$\pm$	0.04	&	1.94	 \\
0.08	&	1.6	&	\nodata	&	3.9	$\pm$	0.04	&	3.64	$\pm$	0.03	&	1.29	 \\
0.08	&	1.7	&	\nodata	&	3.76	$\pm$	0.04	&	3.5	$\pm$	0.04	&	1.54	  \\
0.08	&	1.8	&	\nodata	&	3.7	$\pm$	0.04	&	3.42	$\pm$	0.04	&	1.98	 \\
0.08	&	1.9	&	\nodata	&	3.92	$\pm$	0.05	&	3.63	$\pm$	0.04	&	1.15	  \\
0.09	&	1.0	&	\nodata	&	3.87	$\pm$	0.02	&	3.61	$\pm$	0.03	&	1.34	  \\
0.09	&	1.1	&	\nodata	&	3.92	$\pm$	0.05	&	3.67	$\pm$	0.05	&	1.44	  \\
0.09	&	1.2	&	\nodata	&	3.97	$\pm$	0.04	&	3.71	$\pm$	0.05	&	0.92	  \\
0.09	&	1.3	&	\nodata	&	3.84	$\pm$	0.05	&	3.59	$\pm$	0.05	&	1.63	  \\
0.09	&	1.4	&	\nodata	&	3.94	$\pm$	0.04	&	3.67	$\pm$	0.05	&	1.87	 \\
0.09	&	1.5	&	\nodata	&	3.75	$\pm$	0.04	&	3.49	$\pm$	0.04	&	1.57	  \\
0.09	&	1.6	&	\nodata	&	3.95	$\pm$	0.04	&	3.67	$\pm$	0.03	&	1.44	  \\
0.09	&	1.7	&	\nodata	&	3.79	$\pm$	0.04	&	3.53	$\pm$	0.05	&	1.38	  \\
0.09	&	1.8	&	\nodata	&	3.92	$\pm$	0.02	&	3.63	$\pm$	0.03	&	1.41	  \\
0.09	&	1.9	&	\nodata	&	3.86	$\pm$	0.02	&	3.58	$\pm$	0.03	&	0.993	  \\
0.1	&	1.0	&	\nodata	&	3.87	$\pm$	0.04	&	3.62	$\pm$	0.05	&	1.76	  \\
0.1	&	1.1	&	\nodata	&	3.91	$\pm$	0.02	&	3.66	$\pm$	0.03	&	1.75	 \\
0.1	&	1.2	&	\nodata	&	3.93	$\pm$	0.04	&	3.69	$\pm$	0.05	&	1.42	  \\
0.1	&	1.3	&	\nodata	&	3.85	$\pm$	0.04	&	3.6	$\pm$	0.05	&	1.18	  \\
0.1	&	1.4	&	\nodata	&	3.86	$\pm$	0.04	&	3.6	$\pm$	0.03	&	1.47	  \\
0.1	&	1.5	&	\nodata	&	3.88	$\pm$	0.02	&	3.62	$\pm$	0.03	&	1.25	  \\
0.1	&	1.6	&	\nodata	&	3.85	$\pm$	0.04	&	3.57	$\pm$	0.03	&	1.47	  \\
0.1	&	1.7	&	\nodata	&	3.8	$\pm$	0.04	&	3.53	$\pm$	0.04	&	1.68	  \\
0.1	&	1.8	&	\nodata	&	3.74	$\pm$	0.04	&	3.46	$\pm$	0.03	&	1.38	\\
0.1	&	1.9	&	\nodata	&	3.76	$\pm$	0.05	&	3.49	$\pm$	0.04	&	1.55	  \\
\enddata 
\end{deluxetable*} 

\end{document}